\documentclass[aps,pre,twocolumn,showpacs,amsmath,amssymb]{revtex4}

\usepackage[dvips]{graphicx}
\usepackage{dcolumn}
\usepackage{bm}
\usepackage{color}
\def\be{\begin{equation}}
\def\en{\end{equation}}
\def\bea{\begin{eqnarray}}
\def\ena{\end{eqnarray}}
\def\bec{\begin{equation}\begin{array}{rcl}}
\renewcommand{\theequation}{\arabic{section}.\arabic{equation}}

\def\p{\partial}
\def\ep{\epsilon}
\def\gs{\gtrsim}
\def\ls{\lesssim}

\def\ve{\varepsilon}
\newcommand{\av}[1]{\langle{#1}\rangle}
\newcommand{\AV}[1]{\bigg \langle{#1}\bigg \rangle}
\newcommand{\bi}[1]{\mbox{\boldmath$#1$}}
\newcommand{\pp}[2]{\frac{\partial {#1}}{\partial {#2}}}

\def\hrij{\hat{\bi r}_{ij}}

\def\aQ{\stackrel{\leftrightarrow}{Q}}
\def\aI{\stackrel{\leftrightarrow}{I}}
\def\a1{\stackrel{\leftrightarrow}{1}}

\begin{document}
\title{
Orientational glass in mixtures of elliptic and circular particles:\\ 
Structural heterogeneities, rotational dynamics, and rheology 
}  
\author{Kyohei Takae and Akira Onuki}
\affiliation{Department of Physics, Kyoto University, Kyoto 606-8502, Japan}


\date{\today}

\begin{abstract} 
Using   molecular dynamics simulation 
with an angle-dependent Lennard-Jones potential, we study 
orientational glass with quadrupolar 
symmetry  in   mixtures of  elliptic  particles   
and  circular   impurities in two dimensions. 
With   a mild aspect ratio ($=1.2$) 
and a mild size ratio ($=1.2$), we 
realize a plastic crystal at relatively high 
temperature $T$. With further lowering $T$, 
we find   a structural phase transition 
for very small impurity concentration $c$ and  
pinned disordered orientations for not small  $c$. 
The ellipses are 
anchored by the  impurities in the planar alignment. 
With increasing  $c$,  the orientation domains composed of isosceles triangles 
 gradually become  smaller, resulting in orientational 
glass with  crystal order. In our simulation,    the impurity distribution 
becomes heterogeneous   during    quenching   from liquid,  
which then produces rotational dynamic heterogeneities. 
We also examine rheology in orientational glass 
to predict a shape memory effect and a superelasticity effect, 
where a large fraction  of the strain is due to 
collective orientation  changes.  
\end{abstract}

\pacs{61.72.-y, 61.43.Fs, 64.70.P-, 62.20.fg}


\maketitle


\section{Introduction}

Certain anisotropic molecules  such as KCN,   N$_2$  and ortho-H 
 form a cubic crystal without orientational order 
\cite{ori,Binder,Kob-Binder}. 
 Solids in such a  rotator phase  are often 
 called plastic solids \cite{Sherwood}. 
As the  temperature $T$ is further lowered, 
they undergo   orientational  phase transitions, 
where the crystal structure  changes from a cubic 
 to   noncubic one \cite{St,ori}. 
In  mixtures of anisotropic particles 
such as KCN diluted with KBr, N$_2$ diluted with Ar, 
  and   ortho-H diluted with para-H,    
  the so-called orientational  glass is realized with increasing the 
 impurity concentration $c$  \cite{ori,Binder,Kob-Binder}, 
where  the quadrupolar, orientational degrees of freedom 
are randomly frozen \cite{Sullivan}.  
In such mixtures, a specific-heat peak  \cite{ori,Mertz}  
and a decrease in  one of the  shear  moduli    \cite{ori,sound}  
 have been observed above  the transition for not large  $c$. 
The latter indicates a strong orientation-strain coupling 
\cite{Raedt,Knorr,Bell,Harris}. 
To explain these behaviors, 
molecular dynamics (MD) simulation was also performed 
on mixed cyanides \cite{Klein}. 
It is also remarkable  that 
 one-component systems of 
globular molecules such as ethanol, 
 cyclohexanol, and C$_{60}$ have  rotator phases 
 and  are  orientationally 
arrested  at lower $T$ with weak  specific-heat 
singularities \cite{Seki,Yamamuro,C60}.

 However, not enough  attention has yet 
been paid to  the  physics of orientational glass. 
In contrast, numerous investigations 
 have been made  on   translational glass, 
where positional disorder is frozen   
 \cite{Kob-Binder}. 
In previous MD simulations on diatomic  
 systems \cite{Kob1} and more complex molecular  
systems \cite{Lewis},  the correlations  between the 
translational and rotational degrees of freedom 
have been examined  in glassy states. 
Glassy dynamics was also found 
in monodisperse hard ellipsoids with a slight anisotropy 
\cite{Schilling}. 
  Moreover, in double glass  \cite{Sch},  
 these two kinds of  degrees of freedom 
have  been predicted to freeze at the same temperature.

Recently,  we performed MD  simulation on 
 mixtures of spheroidal particles and spherical 
impurities in three dimensions to examine  
the formation of orientational glass \cite{EPL}. 
In this paper, we  aim  to 
investigate  its complex  dynamics 
 in more detail in   mixtures of  elliptic  particle and 
circular  impurities  in crystal in two dimensions. 
We assume a mild aspect ratio ($=1.2$) 
of the ellipses to avoid liquid crystal mesophases 
 and a mild size ratio  ($=1.2$) 
between the two species  to suppress 
 positional disorder.  We  vary $T$ and  $c$ 
to examine  the changeover between 
multi-variant domain states for  small $c$ 
and highly frustrated states of 
orientational glass for not small $c$.  
We  shall find that  
 mesoscopic orientational order and strains 
exist  in glassy states. 
Previously,  for    binary mixtures of circular particles, 
  the changeover between 
polycrystal and translational glass 
was studied  with  varying  $c$   \cite{Hamanaka}.     In 
translational glass, 
mesoscopic crystalline order still remains  and was visualized 
 \cite{Hamanaka,Tanaka}.  In double glass, 
simultaneous appearance of these  two mesoscopic heterogeneities 
have  been detected \cite{Takae-double}.

If the  molecules forming a  crystal are  anisotropic, 
there arises  a direct  coupling between 
the  orientations and the lattice deformations 
 \cite{Raedt,Knorr,Bell,Harris}. 
In fact,  an effective interaction mediated  by acoustic phonons was derived 
among anisotropic particles in crystal  
such as  (CN)$^-$ in KCN  \cite{Raedt,Knorr,Bell}, 
leading  to acoustic softening in the rotator phase 
 \cite{ori,sound}.  
The orientational  phase transitions for small $c$ thus 
belong to type-I instabilities in   Cowley's   
classification of elastic instabilities  \cite{Cowley,Onukibook},  
  where acoustic modes become soft in particular 
wave vector directions.  
 In this paper, we predict shape memory effect and a superelasticity effect 
 in orientational glass at low $T$, where favored  oriented domains 
increase and disfavored ones decrease 
upon stretching. These  
effects are well-known  for  shape-memory alloys 
such as TiNi \cite{marten,RenReview,Ren}.
Molecular dynamics simulation was 
also performed  to reproduce superelasticity for a model alloy \cite{Suzuki}. 
It is worth noting that mesoscopic strain heterogeneities 
  were observed in TiNi glass, which were 
 on a scale of $20$nm at a slightly 
 off-stoichiometric composition \cite{Ren}.

As another ingredient, 
we shall find a  tendency of impurity clustering 
 depending on the molecular interactions 
\cite{EPL}. In our simulation, it  took  place during 
  quenching from high-temperature liquid to low-temperature solid. 
The impurity clustering 
 gives rise to significant 
heterogeneities in  orientational order 
and  rotational dynamics. 
For example,  a small fraction 
of the  elliptic particles  remain  not strongly anchored 
to  the impurities such that they 
 undergo flip rotations  even at very low $T$. 
Though such   effects have rarely been 
discussed in the literature, 
they should  be relevant 
in many real experiments.

We point out that our system is similar to liquid crystal gels 
(gels containing rodlike molecules)  
 \cite{PGgel,Tere}. In such soft matter, 
 there arises a strong orientation-strain 
coupling, which makes  the isotropic-nematic  transition   
analogous to the orientation transition in solids. 
Irregularities in the  crosslinkage 
play the role of random  quenched disorder,  
leading  to  mesoscopic  nematic polydomains \cite{Uchida}.  
Application of  stress  or 
electric field   induces polydomain-monodomain 
transitions \cite{Kai}. 
We also note that  dilute magnetic alloys, called  spin glass, 
have  glassy phases characterized 
by frozen-in local magnetic 
moments which point in random directions 
\cite{Binder,Sullivan,Kob-Binder}. 
However, in its  research, 
 the spin-lattice coupling 
and the impurity clustering (resulting in  
correlated quenched disorder) 
have  not yet been well examined.

The problem of orientational glass is thus closely 
related to many important 
 problems in solids and soft matter. 
On the basis of  a simple microscopic model, 
we organize  this paper  as follows. 
In Sec.II,   we will 
present  the background of our 
simulation. In Sec.III, we will display 
orientational  configurations 
for various $c$ at low $T$.
In Sec.IV, we will examine 
the rotational  dynamics. 
 In Sec.V, we will treat rheology  
of orientational glass.

\setcounter{equation}{0}
\section{Simulation background}

We use an angle-dependent potential \cite{EPL}, 
which is similar to but much simpler than the  Gay-Berne 
potential   \cite{Gay} for rodlike  molecules forming  
 mesophases.

\subsection{Model of anisotropic particles}
We consider a   binary mixture in two dimensions, 
where the first species  consists of  anisotropic 
particles  with number $N_1$ 
and the second species consists of circular ones 
with   number  $N_2$. 
The total number is  $N=N_1+N_2$. 
The  concentration of the second species is  
\be 
 c=N_2/N.
\en      
For small $c$,   the circular  particles   are  impurities. 
The particle   positions are  
written as ${\bi r}_i$, where $i=1, \cdots, N_1$ 
for the anisotropic particles ($i\in 1$) 
and $i=N_1, \cdots,  N$ 
for the isotropic particles ($i\in 2$). The  
orientations  of the anisotropic particles    
 are expressed by 
\be 
{\bi n}_i=(\cos\theta_i,\sin\theta_i),  
\en 
where $\theta_i$ are their 
 angles $\theta_i$ with respect to the $x$ axis.    
The particle sizes are characterized by two 
lengths, $\sigma_1$ and $\sigma_2$. 
The pair potential $U_{ij}$  between particles 
$i \in \alpha$ and $j\in \beta$ 
($\alpha,\beta=1,2$)  depends on the distance 
 $r_{ij}= |{\bi r}_i -{\bi r}_j|$ and  
the angles   $\theta_i$ ($i\in 1$) and   $\theta_j$ ($j\in 1$).  
For  $r_{ij}< r_c$,  it  is a modified Lennard-Jones potential, 
\be  
U_{ij}
=4\ep\bigg[(1+ A_{ij}) 
\frac{\sigma^{12}_{\alpha\beta}}{r_{ij}^{12}}
- \frac{\sigma_{\alpha\beta}^6}{r_{ij}^6} \bigg] -C_{ij},  
\en 
For  $r_{ij}> r_c=3\sigma_1$, it is zero.   
Here,        $\ep$ is 
the characteristic interaction energy  and 
\be 
\sigma_{\alpha\beta}=
(\sigma_\alpha + \sigma_\beta)/2. 
\en   
The   $C_{ij}$ ensures  the continuity of $U_{ij}$ 
at $r_{ij}= r_c$, so it  is equal to 
the first term  at $r=r_c$ in the right hand side of Eq.(2.3).  
The  angle factor $A_{ij}$  depends  on 
  the  angles  between  
the relative direction 
$\hrij= r_{ij}^{-1}({\bi r}_i -{\bi r}_j)$ 
and the particle orientations. 
In  this paper, we assume the following form,       
\be
A_{ij} =  \chi \delta_{\alpha 1} 
({\bi n}_i\cdot\hrij)^2+ \chi 
\delta_{\beta 1} ({\bi n}_j\cdot\hrij)^2, 
\en 
where   $\delta_{\alpha\beta}$ is the Kronecker delta 
and  $\chi$ represents   the strength 
of anisotropic repulsion for $\chi>0$. 
Our potential is invariant with respect to turnovers  
$\theta_i \to \theta_i\pm \pi$ or inversions ${\bi n}_i \to 
-{\bi n}_i$.  In our system,  it follows  quadrupolar glass    
(orientational glass with quadrupolar symmetry) 
with impurities \cite{Sullivan,Kob-Binder,Binder,EPL}. On the other hand, 
if $U_{ij}$   contains  a term like  
  $({\bi n}_i \cdot{\bi n}_j) v(r_{ij})$, 
it is not invariant with respect to these 
transformations, 
resulting in  spin  glass  with impurities \cite{Binder,Kob-Binder}.

The total kinetic 
energy of our system is   given by   
\be
K =  \frac{1}{2} \sum_{1\le i\le N}{m_\alpha} |{\dot{\bi r}}_i|^2
+  \frac{1}{2} \sum_{1\le i\le N_1} {I_1}|\dot{\theta}_i|^2,
\en  
where ${\dot{\bi r}}_i= d{\bi r}_i/dt$,    
${\dot{\theta}}_i = d\theta_i/dt$,   
$m_1$ and  $m_2$   are 
 the masses,   and $I_1$ is the  moment of 
inertia of the anisotropic particles. The total  potential energy  is 
$U= \sum_{ i<j}U_{ij}$ and the total energy is 
  $H=K+U$. The Newton equations  
 for ${\bi r}_i(t)$ and $\theta_i(t)$ are 
written as 
\bea
&&{m_\alpha} 
{\ddot{\bi r}}_i =\frac{d}{dt}\pp{}{{\dot{\bi r}}_i}K=-\pp{}{{\bi r}_i}U,
\\
&&{I_1} {\ddot{\theta}}_i = 
\frac{d}{dt}\pp{}{{\dot\theta}_i}K= -\pp{}{\theta_i}U,  
\ena 
where    ${\ddot{\bi r}}_i= {d^2}{\bi r}_i/{dt^2}$ 
($i\in 1$ and 2) and  ${\ddot{\theta}}_i= 
{d^2}{\theta}_i/{dt^2}$ ($i \in 1$). 
In our  time integration, 
 $\theta_i$ are unbounded, changing continuously 
 in the range $[-\infty,\infty]$.

We regard  the anisotropic particles 
   as ellipses with  
short and long diameters  $a_s$ and $a_\ell$. 
To determine them, we minimize    
$U_{ij}$ in Eq.(2.3) with respect to $r_{ij}$ 
to obtain   $r_{ij}= 2^{1/6}(1+A_{ij})^{1/6}\sigma_1$. 
For $\chi>0$,  this distance is minimum at $A_{ij}=0$ 
 for the perpendicular orientations  
 (${\bi n}_i, {\bi n}_j \perp  {\hat{\bi r}}_{ij} $) 
and is  maximum at $A_{ij}=2\chi$ 
for  the parallel orientations  
 (${\bi n}_i, {\bi n}_j \parallel   {\hat{\bi r}}_{ij} $). Thus, we set        
\be 
a_s= 2^{1/6}\sigma_{1},\quad  a_\ell=
 (1+2\chi )^{1/6}a_s. 
\en 
If these  elliptic diameters are assumed, we obtain    
\be 
I_1= (a_s^2+ a_\ell^2)  m_1  /4.
\en

\subsection{Simulation  method} 

We integrated  Eqs.(2.7) and (2.8) using the leap-frog method   
under the periodic boundary condition for   $N=N_1+N_2=4096$. 
 We  set  
\be  
 \chi=1.2 , \quad \sigma_2/\sigma_1=1.2.
\en  
The aspect ratio  is   $a_\ell/a_s= 1.23$ from Eq.(2.9). 
We   measure space  in units of $\sigma_1$ and time in units of   
\be 
\tau_0 = \sigma_1 \sqrt{m_1/\ep}. 
\en 
 We also set  $m_1=m_2$. 
The  temperature $T$  is   measured 
in  units of $\epsilon/k_B$ with $k_B$ being the Boltzmann constant.

To  prepare the initial states in each simulation run,   
we started   with a liquid state at $T=2$,   lowered $T$ to 0.5 
below  the melting temperature ($\sim 1)$,  
 waited for a time interval of $10^5$, 
and changed $T$ to the final temperature, 
where we   attached      a Nos$\acute{\rm e}$-Hoover thermostat 
\cite{nose} to all the particles. 
After this preparation, we used  three  simulation methods.
First,  retaining  the thermostat, we took data 
  in the $NVT$ ensemble (Sec.III).
Second,  we switched off  
the thermostat,   waited for another time interval of 
$ 10^5$, and  calculated   the time 
correlation functions  in the NVE ensemble  (Sec.IV).  
In these simulations at fixed  $V$,  
the cell  volume $V$ was given   by    
\be 
 \frac{\pi}{4}  a_sa_\ell (1-c)+
\frac{\pi}{4} a_s^2 c  =0.95\frac{V }{N}. 
\en  
Then the  cell length was      $V^{1/2}\sim  70$ for small $c$.
Third,  to    apply  uniaxial stress, we 
varied the cell  lengths in the $x$ and $y$ 
axes  assuming  a rectangular cell, 
 where we used  the method of 
Parrinello and Rahman \cite{Rahman} (Sec.V).

\begin{figure}[t]
\begin{center}
\includegraphics[width=230pt]{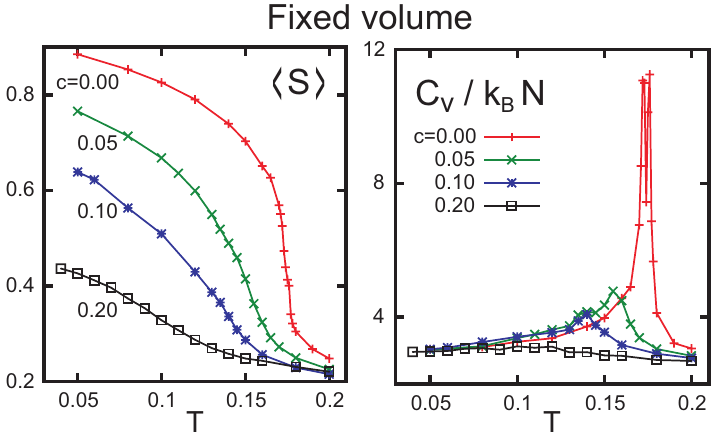}
\caption{Average orientation amplitude 
$\av{S}$  in Eq.(A4) (left) 
and specific heat $C_V$ in Eq.(3.1) 
divided by $k_BN$  (right) 
vs $T$ for  $c=0.0,0.05, 0.1$, and 0.2, 
which are calculated   in  the $NVT$ ensemble. 
The peak of $C_V$ decreases with increasing $c$. 
}
\end{center}
\end{figure}

\begin{figure}
\begin{center}
\includegraphics[width=240pt]{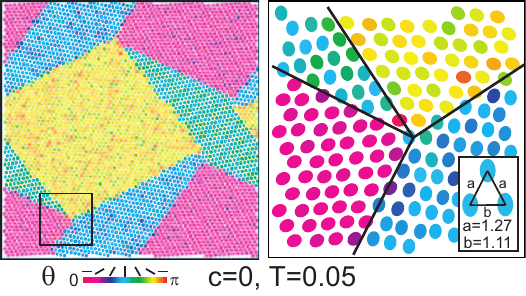}
\caption{Left: Domain pattern 
composed of three martensitic variants for $c=0$ 
at  $T=0.05$. Right: Elliptic particles 
around  a junction  of four  domains 
 in the box region in the left. 
The angles among  the four  lines 
are nearly equal to  $\pi/2$ (for yellow square), 
$\pi/6$, $\pi/2$, and $5\pi/6$, being multiples of $\pi/6$. 
Each variant is composed of isosceles triangles 
 with long side length  
$a=1.27\sigma_1$ and short one $b=1.11\sigma_1$  (inset). 
Colors represent $[\theta_i]_\pi$  according to the color bar.}
\end{center}
\end{figure}

\begin{figure}[t]
\begin{center}
\includegraphics[width=230pt]{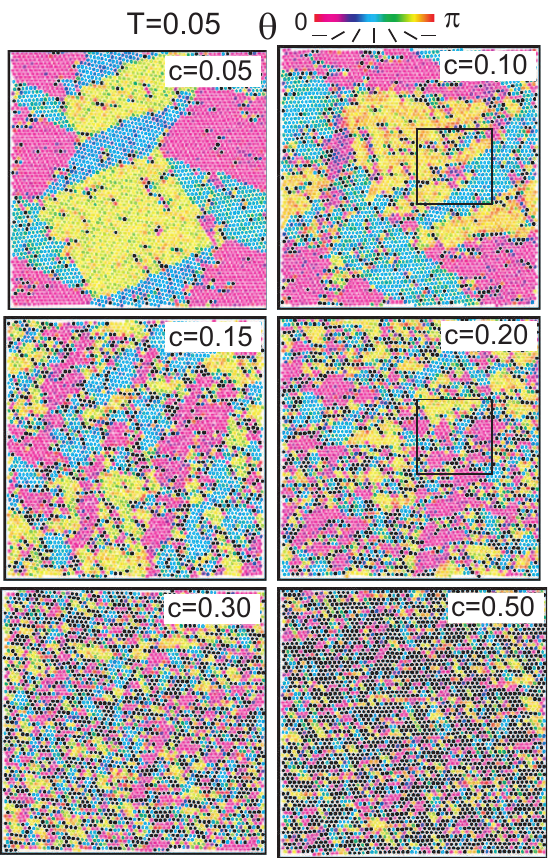}
\caption{Frozen patterns of  the angles $\theta_i$ of the ellipses   
 at  $T=0.05$ for  $c= 0.05, 0.1$, 0.15, 0.2, 0.3, and 0.5. 
Colors of the ellipses represent $[\theta_i]_\pi$  
according to the color bar, while black points ($\bullet$) 
represent impurities. With increasing impurity concentration $c$, 
the domains become smaller, resulting in 
orientational glass.  
}
\end{center}
\end{figure}
 \begin{figure}[t]
\begin{center}
\includegraphics[width=240pt]{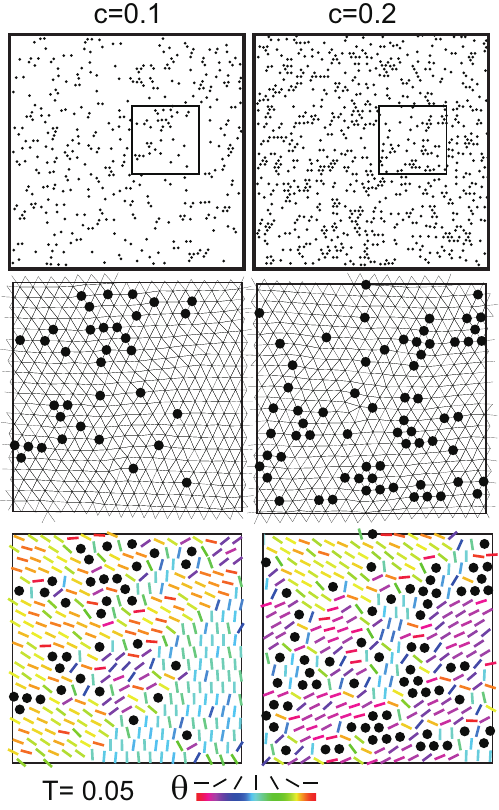}
\caption{Top: Impurity distribution   with significant  clustering 
 for $c=0.1$ (left) and 0.2 (right) at $T=0.05$.  
The data of these snapshots are  common to those 
  of $c=0.1$ and 0.2 in Fig.3. Middle:  Delaunay diagrams 
in  the box regions in the top panels (common to those in  Fig.3), 
mostly composed of isosceles triangles. 
Impurities are also written  $(\bullet)$.   
Bottom: Expanded snapshots of  ellipses  
around  impurities  in the same regions, 
exhibiting  planar  anchoring with $[\theta_i]_\pi$ 
 according to the color bar. 
}
\end{center}
\end{figure}

\setcounter{equation}{0}
\section{{NVT  simulation} of orientational glass on a hexagonal lattice} 
In Figs.1-4, we  give  results   in the  $NVT$ ensemble.
Under Eqs.(2.11) and (2.13),  
the  melting temperature $T_m$ was about $1.0$, 
above which liquid was   realized.  
Below $T_m$,  a hexagonal crystal 
without orientational order  appeared.   
For    $T \ls  0.4$, 
 an orientational phase transition took  place  
 for  small $c \ls 0.1$, 
while orientational 
glass emerged  for not  small $c \gs 0.2$.
Previously,    many authors 
  \cite{Frenkel1}  
 examined the  phase transition 
between  orientationally disordered and 
ordered  crystal phases for  monodisperse hard rods  
with  mild aspect ratios.

\subsection{Order parameter amplitude and specific heat}

In Fig.1, we plot the average orientation amplitude 
$\av{S}$  defined in Appendix A 
and  the constant-volume specific heat $C_V=
( \p {\av{H}}/\p{T})_{NV}$ vs $T$  for four concentrations. 
The former represents the overall strength of the 
orientational order, while  the latter is   the fluctuation amplitude 
of the energy  $H$ in the $NVT$ ensemble,    
\be 
C_V= \av{( H -\av{H})^2}/k_BT^2 .   
\en  
 Hereafter,  $\av{\cdots}$ denotes the average 
over time and over  several runs.  For $c=0$,  $\av{S}$ grows 
nearly discontinuously at $T=0.18$ 
in a narrow temperature window 
with width about $0.02$, where the disordered and ordered phases coexist. 
However, its $T$ dependence becomes gradual for $c>0$.  
The   $C_V$   exhibits a peak at the orientational transition and 
 its   peak height decreases with increasing $c$. 
This  peak stems  from  the enhanced  orientation fluctuations near the 
transition.

In the same situation with the same $N$, 
we also performed simulation in the 
$NpT$ ensemble with an isotropic applied stress 
in two dimensions (not shown in this paper), 
where we allowed the cell to take a rectangular shape. 
There,  the transition was  first-order 
with  discontinuous changes in volume and entropy \cite{comment1}
as in three-dimensional KCN 
 \cite{Mertz,St,Raedt,Bell,Kob-Binder}.
We also found a sharp peak in  the isobaric specific heat 
$C_p$  at the transition. 
For $c\gs 0.1$, the impurities pin 
the domain growth and   the $NVT$ and $NpT$ simulations 
provide  essentially the same  low-$T$  behavior.

In previous experiments on (KCN)$_{1-c}$(KBr)$_{c}$, 
 $C_p$    exhibited  peaks as a function of $T$ 
at structural phase transitions for small $c$,  
but it varied  continuously  without peaks  
for large  $c$  \cite{ori,Mertz}. 
In addition, the peak height for small $c$ was  
of order $5k_B$ per molecule. 
These features are common to 
those of our specific heat  results.

\subsection{Structural phase transition for $c=0$  
and fragmentation of  domains for $c>0$}

In the one-component case $c=0$, 
the ellipses 
undergo a first-order structural  (martensitic) phase transition 
from  a hexagonal lattice 
to a deformed hexagonal lattice formed by  isosceles triangles. 
The  transition temperature is about 0.18 under Eq.(3.1) 
as indicated by Fig.1. 
In Fig.2, we show a typical orientational pattern 
of oriented  domains for  $c=0$ below the transition at fixed $V$. 
Depicted are the angles,
\be 
[\theta_i]_\pi=\theta_i-k\pi
\en  
in the range $[0,\pi]$ with $k$ being an integer, 
where $\theta_i$ and $\theta_i\pm \pi$ are not differentiated. 
There appear  three crystalline variants   
with the same volume fraction $1/3$. 
Here,  ${\bi n}_i$ are aligned along 
one of the underlying crystal  axes, so   
each variant is composed of  isosceles triangles 
elongated along its orientated direction. 
 The  domains  are separated by 
sharp  interfaces, where the surface tension is  
about  $ 0.2$ in units of $\epsilon/\sigma_1^2$. 
As a unique feature, the junctions,  at which  
domain boundaries intersect, 
have angles   $\pi n/6$ ($n=1,2,\cdots)$ approximately.  
Similar  unique  domain 
patterns were experimentally 
observed on hexagonal planes after structural 
phase transitions \cite{Kitano}. 
They were also  reproduced  in 2D   phase-field 
simulations \cite{Chen,Onukibook}.

Next, it is of great interest how the domain structure is 
influenced by impurities. 
In Fig.3,  we present  
snapshots of $[\theta_i]_\pi$     at    $T=0.05$ 
 for  $c= 0.05, 0.1$, 0.15, 
0.2, 0.3, and 0.5. 
At this low $T$,  the thermal fluctuations 
are very small and   the 
 patterns of $[\theta_i]_\pi$ are nearly frozen in time 
 even on a time scale of $10^5$, though flip rotations 
are still activated infrequently (see the bottom panels 
of  Fig.5 below).  We can see that 
the domains gradually become finely divided  with increasing $c$. 
The fractions of the three variants are 
 the same ($1/3)$ at fixed $V$. 
For $c\gs 0.2$, the  orientational disorder 
is much enhanced, resulting in orientational glass.   
 We  note that the  distribution of 
$[\theta_i]_\pi$ is   peaked at the three angles 
along the  crystal directions (even for $c=0.5)$       
 (see Fig.15 below).

\subsection{Impurity clustering and planar anchoring  }

The top panels of Fig.4 display  the   overall 
impurity distributions  for $c=0.1$ and 0.2 at $T=0.05$.  
We can see a significant tendency of 
   clustering of the impurities, which took 
 place mostly during liquid states in the quenching 
process.   In the present model, 
 association  of the impurities 
lowers the total potential energy  
by about  $-5\epsilon$ per impurity \cite{EPL}.
In contrast, such impurity clustering has been neglected 
in spin glass theories \cite{Binder,Kob-Binder}. 

In the middle  plates of Fig.4,  
the Delaunay triangulation 
is given for  the particle configurations in the box regions 
in the upper panels, 
which are the dual graphs of the Voronoi diagrams.  
They are mostly composed of isosceles triangles in the inset of Fig.2. 
With  impurities of size ratio 1.2, the hexagonal lattice  
is locally elongated,  but its structure is preserved, where 
 the number of triangles surrounding 
each particle (the coordination number) $k$  is mostly 6. 
In Figs.3 and 4, there appear  particles  
 with $k=5$ and those with  $k=7$.  They are both two ellipses for $c=0.1$, 
while  they  are both seven 
 (including  two impurities with  $k=7$)  for $c=0.2$. 
See Fig.11(b) below  for such defects. 

In the bottom panels of Fig.4, we present   expanded snapshots 
 of anisotropic  particles around  impurities. The   alignments are mostly 
perpendicular  to the surface normals of the impurities, 
 analogously  to  the  planar  anchoring 
of liquid crystal molecules near colloid 
surfaces  \cite{PG}. Domain boundaries between 
different variants can also be seen.

 To examine the degree of clustering of the  impurities, 
 let us  group them  into clusters. 
 We assume that   two  impurities $i$ and $j$ 
 belong  to  the same cluster 
if their distance is shorter than 
$1.6$. Then, we calculate  
the   numbers $N_{\rm I}(\ell)$ of 
the $\ell$ clusters 
consisting of $\ell$  impurities, where 
$\sum_{\ell \ge 1} \ell  N_{\rm I}(\ell)=N_2$. 
The  average cluster size is given by   
\be 
{ \ell}_{\rm I}= 
\sum_{\ell} \ell^2   N_{\rm I}(\ell)/N_2
\en 
In Fig.3, we have 
 ${ \ell}_{\rm I}=1.37$, 2.04, 3.02, 
 4.79, 10.7, and 984  
  for $c=0.05$, 0.1, 0.15,  $0.2$, 0.3, and 0.5, respectively. 
For $c=0.5$,  a  cluster of the system size 
$\sim 1400$ appears.


 \setcounter{equation}{0}
\section{{NVE  simulation} of rotational dynamics }

In Figs.5-12, we  give simulation results   in the  $NVE$ ensemble, 
 where  the average translational and rotational  kinetic energies   
were  kept at $k_BT$ and $k_BT/2$, respectively,  per particle. 
 Varying $T$, we  
examine the rotational dynamics  at  $c=0.2$.  
For $0.3\ls T\ls 0.7$,  we realize the rotator phase, where 
 non-flip rotations with angle changes not close to $\pm \pi$ 
 are gradually arrested with lowering  $T$. 
For $T\ls 0.3$,  quadrupolar  glass is realized, where  
  only flip rotations can be activated heterogeneously.

\begin{figure}[t]
\begin{center}
\includegraphics[width=240pt]{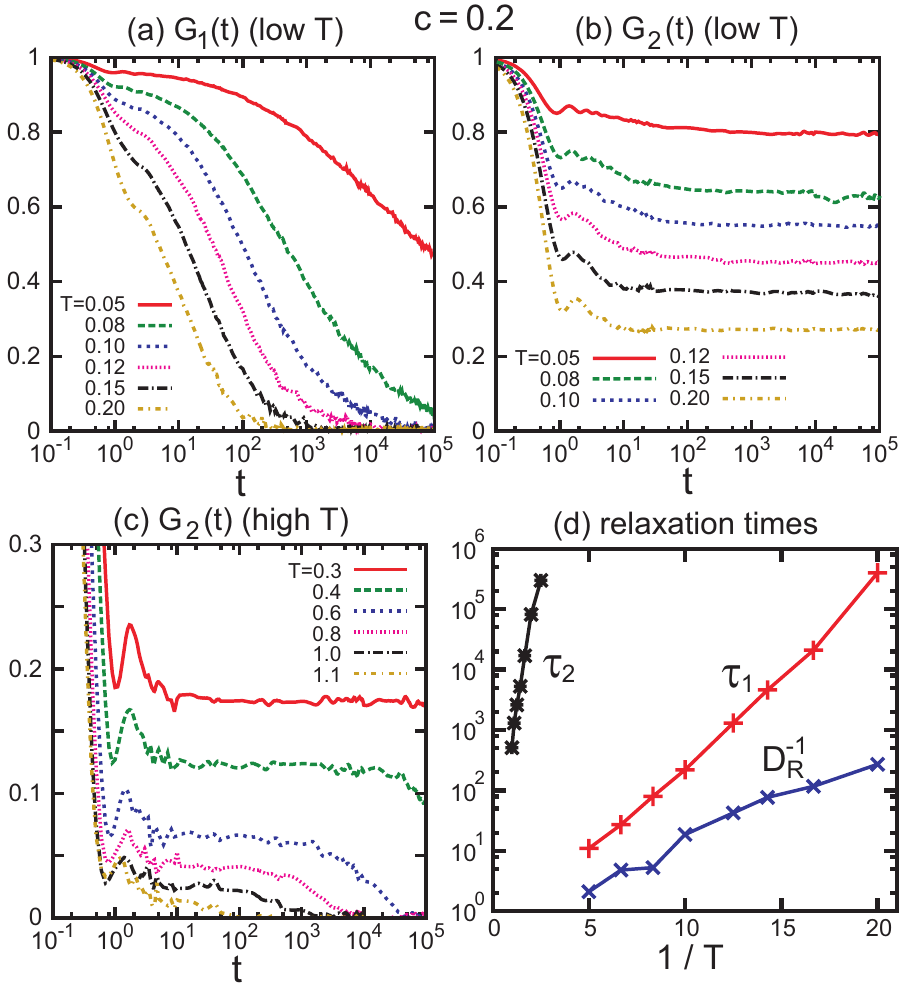}
\caption{Orientational  relaxations for 
$c=0.2$ at  various  $T$. (a) $G_1(t)$  and (b) $G_2(t)$   
 at low $T=$0.05,0.08,0.10,0.12, 0.15, and 0.20.    
(c) $G_2(t)$   
 at relatively high $T=$0.3,0.4,0.6,0.8, 1.0, and 1.1.
(d) Relaxation times $\tau_1$ from $G_1(t)$ 
at low $T$ in Eq.(4.3),  $\tau_2$ 
from $G_2(t)$ at high $T$ in Eq.(4.4),  
and the inverse  rotational diffusion 
constant $D_R^{-1}$ in Eq.(4.12). 
 }
\end{center}
\end{figure}
 
\begin{figure}[t]
\begin{center}
\includegraphics[width=240pt]{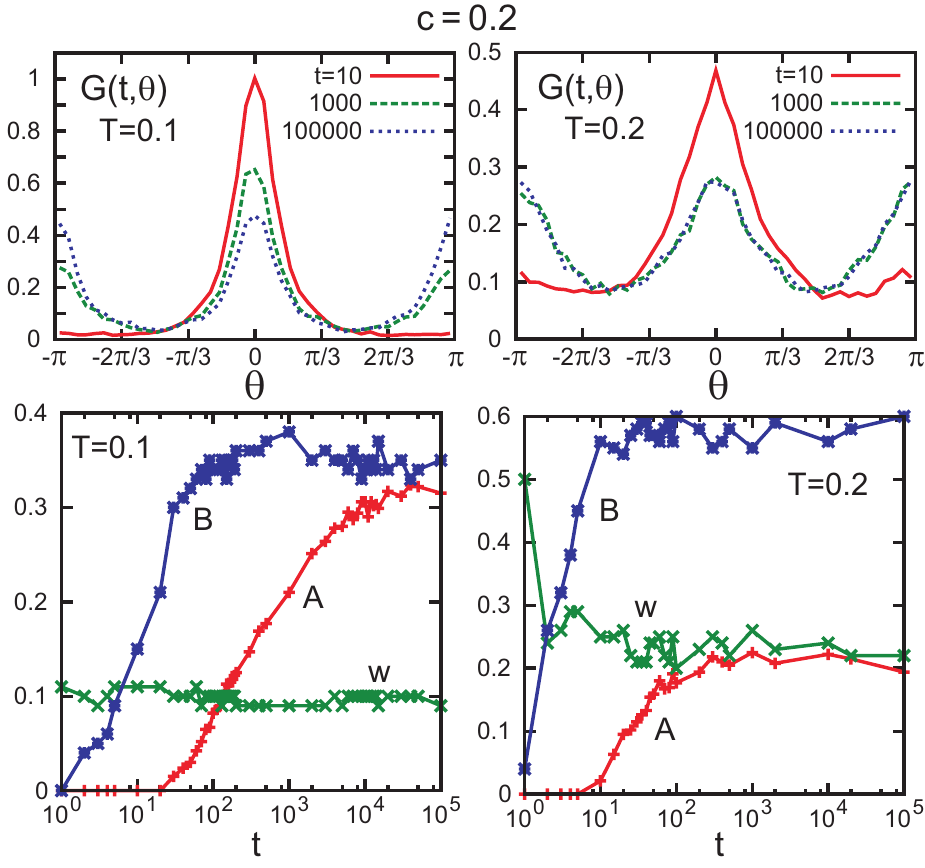}
\caption{Top:Time-dependent angle distribution  
$G(t,\theta)$ in Eq.(4.5) for  $c=0.2$  at $t=10, 10^3$, and $10^5$. 
Here,  $T=0.1$ and $\tau_1=220$ (left), while  $T=02$ and  
$\tau_1=220$ (right).
Bottom: Parameters $A$, $B$, and $w$ in the approximate expression 
(4.8) for  $T=0.1$  and 0.2  as functions of $t$. 
}
\end{center}
\end{figure}  

\begin{figure}[t]
\begin{center}
\includegraphics[width=245pt]{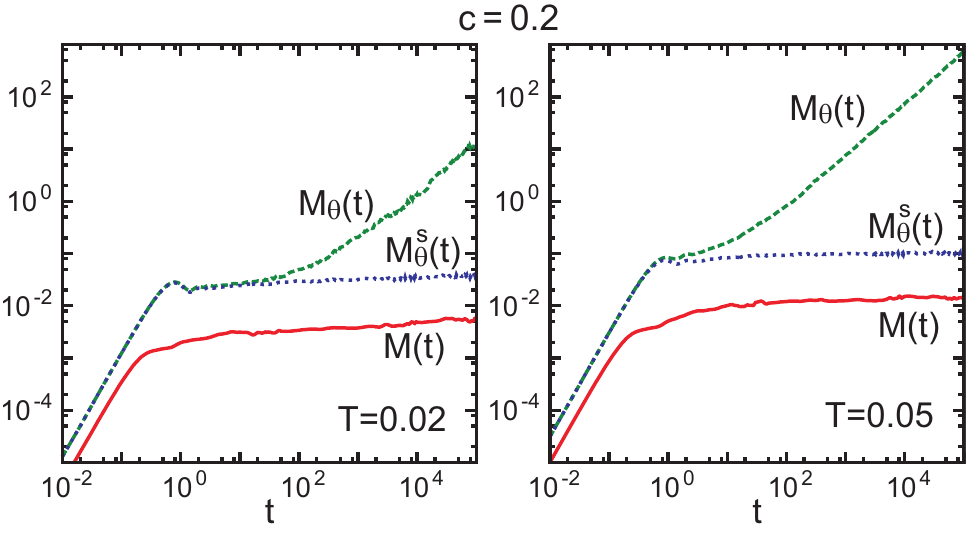}
\caption{
$M_\theta (t)$  in Eq.(4.11),  $M (t)$  
in Eq.(4.13), and $M_\theta^s(t)$ in Eq.(4.14) 
for $c=0.2$ at $T=0.02$ (left) 
and $T=0.05$ (right). These  are  the averages of 
 $(\Delta\theta_i)^2$,  $|\Delta {\bi r}_i|^2$, and 
 $\sin^2(\Delta\theta_i)$, respectively.  
 At these low $T$,   $M (t)$   and $M_\theta^s(t)$ tend to 
plateaus, but $M_\theta(t)$ 
exhibits linear growth at long times  due to flip rotations.   
Then, $D_R$ is $0.75\times 10^{-4}$ at $ T=0.02$ 
and  $3.5 \times 10^{-3}$ at $T=0.05$. 
Transient plateau behavior can also be seen 
in $M_\theta(t)$ at $T=0.02$. }
\end{center}
\end{figure} 

\subsection{Rotational time-correlation functions}

The rotational dynamics   has  been 
extensively investigated for 
anisotropic particles  in glassy states \cite{Kob1,Lewis,EPL,Schilling,Sch}. 
We consider  the 
 rotational time-correlation functions 
 $G_1(t)$ and $G_2(t)$ for the ellipses  defined by 
\be
G_\ell (t)=     \frac{1}{N_1} \sum_{j\in 1} 
\AV{\cos[\ell\Delta \theta_j(t_0, t_0+t)]}   , 
\en 
where $\ell=1$, 2. We write the angle change of 
ellipse $i$ as    
\be 
\Delta\theta_i(t_0,t_1)=\theta_i(t_1)-\theta_i(t_0).   
\en   
In 3D,  use has been made 
of the Legendre polynomials $P_\ell(\theta)$.   
In  Fig.5, we show 
  $G_1(t)$ and  $G_2(t)$  for  $c=0.2$ at various  temperatures.  
For low $T \le  0.2$, 
the relaxation of $G_1(t)$ 
is slowed down with lowering  $T$ in (a), while  
 $G_2(t)$ tends to a plateau $f_2(T)$ 
for $t\gs 10^2$  after considerable  initial  relaxations  
in (b). For higher $T \ge 0.4$, 
 $G_2(t)$ decays at long times in  (c). Relaxation times vs $1/T$ are 
plotted in (d), where  $\tau_1$  of $G_1(t)$ 
is determined by   
\be 
G_1(\tau_1) = e^{-1}. 
\en  
At long times,   $G_2(t)$  relaxes only 
for $T\ge 0.4$  and its relaxation time  $\tau_2$  
is determined by its fitting to 
the  stretched exponential form, 
\be 
G_2(t) \cong   f_2 \exp[-(t/\tau_2)^{\beta}]. 
\en  
In our case, we find $\beta\cong 1$. 
In the  Arrhenius form,  
we obtain $\ln(\tau_1)=0.68/T-1.2 $ for $T\ls 0.2$ and 
$\ln(\tau_2)=4.2/T+2.5$ for $T\gs 0.4$. Note that  these two 
temperature ranges are separated.

The difference between $G_1(t)$ and $G_2(t)$ 
can  be understood if we consider the  distribution 
of the angle changes,   
\be 
G(t,\theta)
=\frac{1}{N_1}\sum_{ i \in 1}
 \AV{\delta([\Delta\theta_i(t_0,t+t_0)]_{2\pi}-\theta)} ,
\en 
where  $-\pi \le \theta<\pi$. 
 For any $\Delta\theta_i$,  we set   
\be 
[\Delta\theta_i]_{2\pi}=\Delta\theta_i-2k\pi, 
\en  
in the range $[-\pi,\pi]$ with  an integer $k$.  
This     $G(t, \theta)$ tends to $\delta(\theta)$ as $t \to 0$  and 
broadens  gradually  for $t>0$. 
 The $G_\ell(t)$ in Eq.(4.1) can be  written  as  
\be 
G_\ell(t)= \int_{-\pi}^{\pi}d\theta 
G(t,\theta) \cos(\ell\theta).
\en
 
In the top panels of Fig.6, we plot time-evolution of 
$G(t,\theta)$ for $c=0.2$ at $T=0.1$ and 0.2. Salient features are as follows. 
(i) The width of the  peak at $\theta=0$,  written as $\sqrt{w}$, 
 soon tends to be  
 independent of $t$. For $t \gg 1$, 
 $w$ represents the   vibrational  amplitude  
of the ellipses (see Eq.(4.10)).  We obtain   $w= 0.09$,  0.17,  and 0.22 
at $T=0.1$, 0.15, and $0.2$, respectively. 
(ii) For $t\gs \tau_1$,   $G(t, \theta)$ 
exhibits  secondary  peaks  at $\theta = \pm \pi$ 
 due to the flip  motions $\theta_i \to \theta_i\pm \pi$.
Their peak widths are  nearly equal to that of the main peak at $\theta=0$. 
(iii) The midpoint values   $G(t, \pm \pi/2)$   become appreciable 
at  long times.

  We  may thus approximate $G(t, \theta)$  as  
a superposition of a constant and  Gaussian functions as
\bea
&& \hspace{-1cm}
G(t, \theta) 
\cong  \frac{B}{2\pi}  +\frac{1-A-B}{\sqrt{2\pi w}}
{ e^{-{\theta^2}/{2w}}}
\nonumber\\
&&\hspace{-1.2cm} +  \frac{A}{\sqrt{2\pi w}} 
\bigg[{e^{-{(\theta-\pi)^2}/{2w}}}
+{e^{-{(\theta+\pi)^2}/{2w}}}\bigg]  .   
\ena  
where $A$ is the turnover probability and 
  $B/2\pi$ is the homogeneous part. 
We fit  the calculated $G(t,\theta)$ to the above form 
to obtain $w(t)$,  $A(t)$, and $B(t)$  vs  $t$ 
for $T=0.1$ and 0.2 in the bottom  panels of Fig.6. 
Here,  $w(t)$ and $B(t)$ are  nearly constant for 
$t \gg 1$, while   $A(t)$  tends to saturate for 
$t\gs 10^5$ at $T=0.1$ and for $t \gs 10^3$ at  $T=0.2$. 
Also  $w$ remains  so small such that  the Gaussian functions in Eq.(4.8) 
are negligible  at the midpoints $\theta=\pm \pi$ 
 compared to   $B/2\pi$.  The plateau  $f_2$ of 
$G_2(t)$  in Fig.5(b) is expressed   as   
\be 
 f_2 \cong (1-B) \exp({-2w}).
\en 
Substitution of  the calculated values of  $B$ and $w$ 
into the above expression yields 
$f_2=0.54$, 0.37, and 0.26 
for $T=0.1$, 0.15, and 0.2, respectively, 
in excellent  agreement  with $f_2$ in  Fig,5(b). 
On the other hand,  at  higher $T (\gs 0.4)$, 
the system is in the plastic solid phase 
 and  $G(t, \theta) $  tends  to be homogeneous   
($=1/2\pi$) very slowly for $t \gg \tau_2$, 
leading to the long-time decay of  $G_2(t)$ in Fig.5(c).

We  comment on  the meaning of $w$ in Eq.(4.8). 
 Let $\bar{\theta}_i$ be the  
time average of $\theta_i$ over many vibrations, 
where we neglect flip rotations. 
Then, we have $\Delta{\theta}_i(t_0,t_1)= \delta\theta_i(t_1)- 
\delta\theta_i(t_0)$ in Eq.(4.2), 
where  $\delta{\theta}_i= {\theta}_i- \bar{\theta}_i$ is 
the deviation from the equilibrium angle  $\bar{\theta}_i$.  
With increasing  $t= t_1-t_0$,  
  $\delta\theta_i(t_1)$ and $\delta\theta_i(t_0)$ 
should become   uncorrelated to give 
$\sum_i \delta\theta_i(t_1) 
\delta\theta_i(t_0)/N_1=0$ so that 
\be 
w= 2\av{\delta\theta^2}=2\sum_{i\in 1}
|\delta\theta_i|^2/N_1,
\en 
where $\av{\delta\theta^2}$ is the variance 
of $\delta\theta_i$ over all the ellipses.

\subsection{ Angular  mean-square  displacement}

In the literature \cite{Kob1,Lewis},  the rotational diffusion has 
been discussed  in terms of 
the angular  mean-square  displacement of the ellipses, 
\be 
M_\theta(t)= \av{(\Delta\theta)^2}= \frac{1}{N_1} 
\sum_{i \in 1} \av{ [\Delta\theta_i(t_0,t_0+t)]^2}.
\en 
which exhibits the ballistic behavior $(\propto t^2)$ 
for $t\ls 1$  and the diffusion behavior for $t\gg 1$ as  
\be 
M_\theta(t) \cong 2D_Rt .
\en 
See Fig.5(d), where  $\ln(D_R^{-1})=0.33/T-0.68$ in the  Arrhenius form.  
 These behaviors are analogous 
to those of  the positional 
mean-square displacement, 
\be 
M(t)= \av{|\Delta{\bi r}|^2}= \frac{1}{N_1} 
\sum_{i\in 1}\av{  |\Delta{\bi r}_i(t_0,t_0+t)|^2}. 
\en 
where $\Delta{\bi r}_i(t_0,t_0+t)=
{\bi r}_i(t_0+t)-{\bi r}_i(t_0)$ 
is the displacement vector of ellipse $i$ in time interval $[t_0, t_0+t]$. 
In Fig.7,  we plot  $M_\theta(t)$ 
and  $M(t)$ vs $t$ at low $T=0.02$ and 0.05 for   $c=0.2$. 
Here, $M(t)$ 
saturates at  a plateau, but  $M_\theta(t)$ still  
exhibits   the diffusion  behavior. 
As Fig.6 suggests, this difference originates  from 
flip rotations without positional changes. 
To confirm this, we also plot 
 the mean-square displacement  
of  $\sin(\Delta\theta_i)$ written as 
\be 
M_\theta^s(t)
=\frac{1}{N_1} 
\sum_{i \in 1}  |\sin[\Delta\theta_i(t_0,t_0+t)]|^2/N_1, 
\en 
which is insensitive to  flip rotations. 
As ought to be the case, 
$M_\theta^s(t)$ coincides with $M_\theta(t)$ at short times 
but  tends  to a plateau at long times.
We also notice  that $M_\theta(t)$ exhibits a plateau in the 
range $1\ls t\ls 100$ at $T=0.02$.


 \begin{figure}[t]
\begin{center}
\includegraphics[width=240pt]{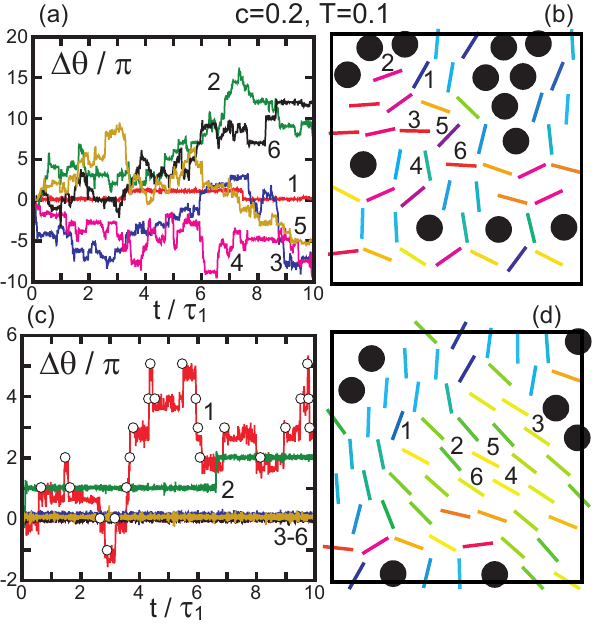}
\caption{Time-evolution of 
 angle changes $\Delta\theta_i(t_0,t_0+t)$ 
in Eq.(4.1)  (left)  for $c=0.2$ and $T=0.1$, where 
$\tau_{1}=220$. Selected  ellipses are 
numbered $1,\cdots ,6$ and  impurities are 
written as large circles  ($\bullet$) (right). 
(a) Rotationally 
active ellipses not anchored by   impurities. 
Ellipse 2 is active though it is rather close 
to impurities.    
(c)  Inactive ones  in an  
orientationally ordered domain ($3-6$) 
and relatively active ones in an interfacial region 
($1$ and $2$).     On the curve 
of ellipse 1, the flip times are marked ($\circ$)  
(see the  Appendix B). }
\end{center}
\end{figure}

 \begin{figure}[t]
\begin{center}
\includegraphics[width=245pt]{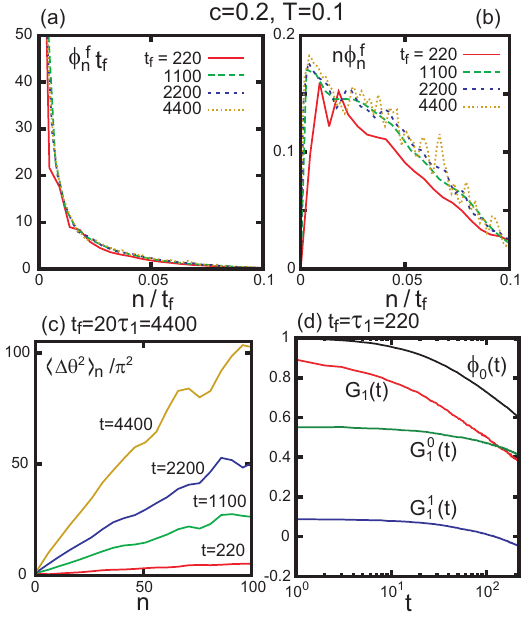}
\caption{(a)  $t_f \phi_n^f$ and  (b) 
$n \phi_n^f$ vs $n/t_f$ for $t_f/\tau_1=1,5,10,$ and 20 
with $\tau_1=220$. (c)  $\av{\Delta\theta^2}_n(t)/\pi^2 $ 
vs $n$  at    $t/\tau_1=1,5, 10$, and 20. Here, 
 $c=0.2$  and $T=0.1$. See Eqs.(4.15) and (4.18) 
for $ \phi_n^f$ and $\av{\Delta\theta^2}_n(t)$.  
(d)  $G_1(t)$ in Eq.(4.1), $G_1^n(t)$ ($n=0,1$) 
with $t_f=\tau_1$ in Eq.(4.23),
 and $\phi_0(t)$ in Eq.(4.24).    
 }
\end{center}
\end{figure}  

 \begin{figure}[t]
\begin{center}
\includegraphics[width=240pt]{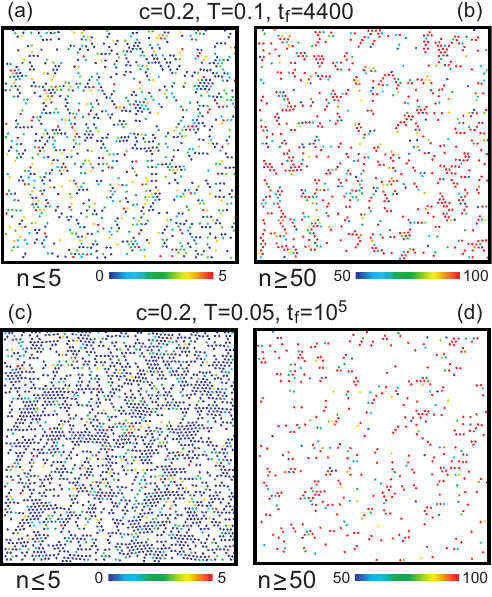}
\caption{Top: Snapshots of ellipses at $T=0.1$ 
with $n_i \le 5$ in (a) 
and   with $n_i\ge 50$ in (b). Bottom: 
Snapshots of ellipses at $T=0.05$ 
with $n_i \le 5$ in (c) 
and   with $n_i\ge 50$ in (d). 
Depicted ellipse fractions  are (a) 0.36, (b) 0.27, (c) 0.72, and (d) 0.15.  
These rotationally inactive and active ellipses  
exhibit heterogeneities closely  
correlated with the impurity clustering (see the 
top right panel in Fig.4).
}
\end{center}
\end{figure}

\subsection{Flip rotations in orientational glass} 

The rotational activity of 
the anisotropic  particles sensitively  
depends on   the surrounding particle configurations. 
In Fig.8, we show typical time-evolution of the 
angle changes. Rotationally inactive 
ellipses  are those anchored to  impurities 
and those within orientationally ordered domains, while  
rotationally active ones  are   those in disordered regions 
not anchored to  impurities and those in interfacial regions 
between different domains.

As will  be shown in Appendix B, we may numerically 
determine  flip events. That is, within  any time interval $[t_0, t_0+t_f]$, 
each ellipse  $i$ flips at successive   times 
$t_0+t_{i1},t_0+t_{i2}, \cdots, t_0+ t_{in_i}$ with 
$t_{i1}>0$ and  $t_{in_i}<t_f$, where  $n_i$ is   
the flip number of ellipse $i$. 
The fraction of  the ellipses with $n$ flips is 
expressed as   
\be 
\phi_n^f  = \av{\sum_{i \in 1}\delta_{n n_i}}/N_1,   
\en 
where  $\sum_{n} \phi_n^f=1$. 
We do not write the $t_f$-dependence 
of $n_i$ and $\phi_n^f$ explicitly. 
We   divide the  ellipses 
into groups ${\cal F}_0$, 
 ${\cal F}_1$, $\cdots, {\cal F}_{n_{\rm max}}$, 
where those  in  ${\cal F}_n$ 
 have undergone $n$ flips in the time interval $[t_0,t_0+t_f]$. 
We introduce the maximum   flip number  $n_{\rm max}$ among all the 
ellipses. For large $n$ and $t_f$, we should have the scaling relation, 
\be 
\phi_n^f= \Phi_f(n/t_f)/t_f,
\en 
 where $\Phi_f(x)$ is a scaling function. In particular, $n_{\rm max}$ 
is proportional to $t_f$ as 
\be 
n_{\rm max}= A_{\rm max}t_f.
\en  
The coefficient  $A_{\rm max}$ is 
about $  0.11$  at  
$T=0.1$ and $   0.022$  at  $T=0.05$. 

 We then introduce 
the angular mean-square displacement within 
the group ${\cal F}_n$   as
 \be 
\av{\Delta\theta^2}_n(t)  = \frac{1}{N_1\phi_n^f} 
\AV{\sum_{i \in {\cal F}_n}  [\Delta\theta_i(t_0,t_0+t)]^2}.
\en 
which is a function of  $t$  for  each given   $t_f$. 
The total angular mean square displacement 
is expressed as  
\be 
M_\theta(t)= \sum_{1\le n\le n_{\rm max}} \phi_n^f 
\av{|\Delta\theta|^2}_n(t)  .
\en 
For sufficiently large $t$ and $n$, 
the ellipses in the group $ {\cal F}_n$ 
should have undergone  $nt/t_f$ flips on the average. 
Then, in the diffusion regime, we should have 
\be 
\av{\Delta\theta^2}_n \cong  \pi^2n t/t_f.  
\en 
Here,  the angle changes 
$\Delta \theta_i(t_0,t_0+t)$ 
at jumps are  $\pi$ or  $-\pi$ and 
their distribution should be  nearly Gaussian.


In Fig.9, we show numerical results 
 which are the averages over  six runs. 
We plot   $t_f \phi_n^f=\Phi_f(x)$ in (a) 
and $n \phi_n^f=x\Phi_f(x)$ in (b) 
 as functions of  $x=n/t_f$  ($n \ge 1$)  
for  $c=0.2$ and  $T=0.1$ by  
setting $t_f/\tau_1=1, 5, 10,$ and  20 ($\tau_1=220)$. 
These curves are nearly independent of $t_f$, 
which confirms the scaling form (4.16). 
However, the ellipses without flips 
  still remain, whose fraction  is 
$\phi_0^f=0.17$ even for $t_f=20\tau_1$. 
From Fig.9(b),  for large $n$, 
 $\phi_n^f$ may be fitted  to  
\be 
\phi_n^f \cong  A_f \bigg( \frac{1}{n} - \frac{1}{n_{\rm max}}\bigg) ,
\en 
where $A_f\cong 0.20$  and 
$n_{\rm max}\cong 500$  at $T=0.1$. 
The  $A_f$ is independent of $t_f$. 
In (c), we also confirm Eq.(4.20). 
From Eqs.(4.12), (4.17), and (4.19)-(4.21), 
  we obtain 
\be 
D_R \cong  \frac{1}{4}  \pi^2 A_f A_{\rm max},
\en   
which yields  $D_R \sim  0.056\sim 13/\tau_1$ at $T=0.1$ 
in good agreement with the  result from the slope 
of $M_\theta(t)$.  At $T=0.05$, 
 we again find Eq.(4.21) with  $A_f  \cong 0.048$,  
 $n_{\rm max}\cong 2800$, and  $\phi_0^f\cong 0.58$ 
for $t_f=10^5$. From Eq.(4.22) these  lead  to 
 $D_R=0.0033\sim 1500/\tau_1$, while  the right panel of Fig.7 
yields $D_R= 0.0035$. 
The rotational diffusion constant $D_R$ 
is thus  determined by the rotationally active 
ellipses with  $n_i \sim n_{\rm max}$.

In contrast, the main contribution 
to $G_1(t)$ in Eq.(4.1) is from the ellipses 
which have  undergone no flip in time 
interval $[t_0,t_0+t]$. This  is the reason why 
 $\tau_1$ behaves very  differently from $D_R^{-1}$ in Fig.5(d). 
  To show this, we set $t_f=\tau_1$. 
We then consider the  following partial sums, 
\be 
G_1^n(t) =  \frac{1}{N_1} \AV{ \sum_{i \in 1, n_i= n } 
\cos[\Delta \theta_i(t_0, t_0+t)]}   . 
\en  
In  Fig.9(d), we  compare 
$G_1^0(t)$ (no-flip contribution) 
and $G_1^1(t)$ (single-flip contribution) with 
 $G_1(t)$. Here, $G_1^0(\tau_1) =0.41=1.11/e$ and 
$G_1^1(\tau_1) =-0.05$, so   $G_1(\tau_1)=1/e$  
mostly consists of the no-flip contribution. 
We also display the fraction of the ellipses 
with no flip in time interval $[t_0,t_0+t]$, 
denoted by  $\phi_0(t)$. Treating  $\phi_0^f$ in Eq.(4.15) 
 as a function of   $t_f$, we have 
\be 
\phi_0(t) =[\phi_0^f]_{t_f=t}. 
\en 
In Fig.9(d), we find   $\phi_0(\tau_1) =0.60$. 
If $\phi_0(t)$ is shifted by 0.1 downward, it   
 nearly coincides with $G_1(t)$.

In Figs.10, we show snapshots of the ellipses 
with $n\le 5$ (left) and $n \ge 50$ (right) for $T=0.1$ and $t_f=4400$ 
(top)  and for 0.05 and $t_f=10^5$  (bottom). The distributions of these 
rotationally inactive and active  ellipses are highly heterogeneous. 
This marked feature is  
due to the significant impurity clustering in the top right 
panel in Fig.4. 
With lowering $T$,  the  flip rotations become increasingly  
infrequent.  In fact, the fraction of the ellipses with $n\ge 50$ 
are 0.36, 0.15, and 0.005  for $(T,t_f)=(0.1,4400)$, 
$(0.05, 10^5)$, and $(0.02, 10^5)$, respectively.

 \begin{figure}[t]
\begin{center}
\includegraphics[width=240pt]{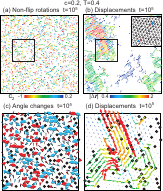}
\caption{ Non-flip rotations and displacements 
between two times $t_0$ and $t_0+t$ with $t=10^5$  
for $c=0.2$  and $T=0.4$, where  orientational glass is approached 
from plastic solid.  Top: (a) Particles with 
 $c_{2i}=\cos[2\Delta\theta_i]<0.2$  forming clusters  
and (b) those with large displacements $\Delta{ r}_i >0.4$ 
caused by defect motions. 
Depicted particle numbers 
are $0.5N_1$ in (a) and $0.35N$ in (b). 
Colors represent $c_{2i}$ and $\Delta{ r}_i$   
according to the color bars.  Expanded snapshot of a defect 
is also given (inset). 
 (c) Expanded snapshot of the ellipses 
in  the box region in (a), 
 where each ellipse $i$ is written as an circular sector 
with two arcs parallel to  ${\bi n}_i(t_0)$ 
and   ${\bi n}_i(t_0+t)$. They are written in blue  for clockwise rotation 
($\Delta\theta_i <0$) 
and in red  for counterclockwise rotation ($\Delta\theta_i >0$). 
Impurities are also depicted($\bullet$).  
(d) Expanded snapshot of the particles with large displacements 
in the box region in (b), where red (green) arrows 
represent displacements by two (one) lattice constants. 
 }
\end{center}
\end{figure}

\subsection{Non-flip rotations in plastic solids 
at relatively high temperature}

So far we have studied  the rotational 
dynamics in orientational glass. We   should also 
examine     the crossover from 
 plastic solid to  orientational glass at higher $T$. 
 In this high $T$-regime,  the long-time decay of 
$G_2(t)$ first saturates at the plateau 
in Eq.(4.9) and slowly decays to zero on the time scale of $\tau_2$  
as in Fig.5(c). Some ellipses are attached to 
impurities on  very long time scales 
and the homogenization of $G(t,\theta)$ takes 
a very long time.

In Fig.11, we hence display  large  non-flip rotations and long-distance 
 displacements  between two times $t_0$ and $t_0+t$ 
with $t=10^5$ for  $c=0.2$ and $T=0.4$. 
Here, the particle positions depicted  are those ${\bi r}_i(t_0)$ 
at the initial time $t_0$.  The flip numbers of the ellipses 
in  this time interval are huge, ranging  from $10^3$ to $10^4$. 
In (a), we pick up 
the ellipses  with    $c_{2i} (t)<0.2$, where we define     
\be 
c_{2i}(t)= \cos[2\Delta{\theta}_i(t_0, t_0+t)]  .
\en 
Because  $c_{2i}$   is invariant with respect to 
turnovers ${\theta}_i \to {\theta}_i\pm \pi$,  
it deviates from unity significantly  due to  non-flip rotations. 
The condition $c_{2i}(t)<0.2$ 
means  $0.22 \pi<[\Delta{\theta}_i(t_0, t_0+t)]_\pi < 0.78 \pi$ 
in terms of $[\Delta\theta_i]_\pi $ in Eq.(3.2).
In (b), we mark  the particles with  $\Delta{ r}_i(t)= 
|{\bi r}_i(t_0+t)-{\bi r}_i(t_0)|>0.4$. These displacements    are 
induced by intermittent 
motions of a few  pointlike defects.  
As in the inset of Fig.11(b), they 
are  composed of two  particles with 
their coordination numbers equal to 
five  and seven (see the explanation of the middle panels 
if Fig.4 also). 
These defects have been observed 
in a number of simulations and experiments in two dimensions  
 \cite{defects}. 
The lengths of the large displacements 
are then mostly  $a$ or $2a$ with $a$ being the lattice constant, 
so they do not affect    the hexagonal crystal structure.
In (c), an expanded  snapshot of the box region in (a) 
is presented, where each ellipse $i$ is written as an circular sector 
with two arcs parallel to ${\bi n}_i(t_0)$ 
and   ${\bi n}_i(t_0+t)$. 
In (d),  an expanded  snapshot of the box region in (b) 
is presented with displacement vectors in arrows.

In Fig.11(a), we can see 
marked  clustering of many  ellipses with 
significant non-flip rotations, which is 
strongly  correlated with the heterogeneous impurity 
distribution. In Fig.11(c), we further notice 
the presence of considerable thermal motions superimposed. 
Also in regions without defects 
(in the upper middle part from Fig.11(b)), 
we may also write   expanded 
figures, but they are  similar to Fig.11(c). 
 Let  $N_{\rm cl}(\ell)$ be the   numbers of 
the $\ell$ clusters consisting of $\ell$  ellipses 
with $c_{2i}<0.2$, where two ellipses $i$ and $j$  belong to the same 
cluster for  $r_{ij}< 1.6$. 
As in Eq.(3.3), the  average cluster size 
may be defined as   
\be 
{ \ell}_{\rm cl}= 
\sum_{\ell} \ell^2   N_{\rm im}(\ell)/N_1.
\en 
Then we find  ${ \ell}_{\rm cl}=48$ for the snapshot in Fig.11(a).

\setcounter{equation}{0}
\section{Rheology  in 
orientational glass}
In Figs.12-15,  we  imposed  
 a Parrinello-Rahman barostat \cite{Rahman} 
  together with a Nos$\acute{\rm e}$-Hoover thermostat \cite{nose}  
 under the periodic boundary condition.  
In our system, small crystalline  domains 
 are elongated along the orientations 
of the ellipses and   
their orientation changes can give rise to 
a macroscopic strain.  
  We here  predict  a shape memory effect 
in orientational  glass, where  
 soft elasticity appears   without  dislocation formation. 
 See Ref.\cite{EPL} for  preliminary results of 3D simulation 
on rheology.

\begin{figure}[t]
\begin{center}
\includegraphics[width=245pt]{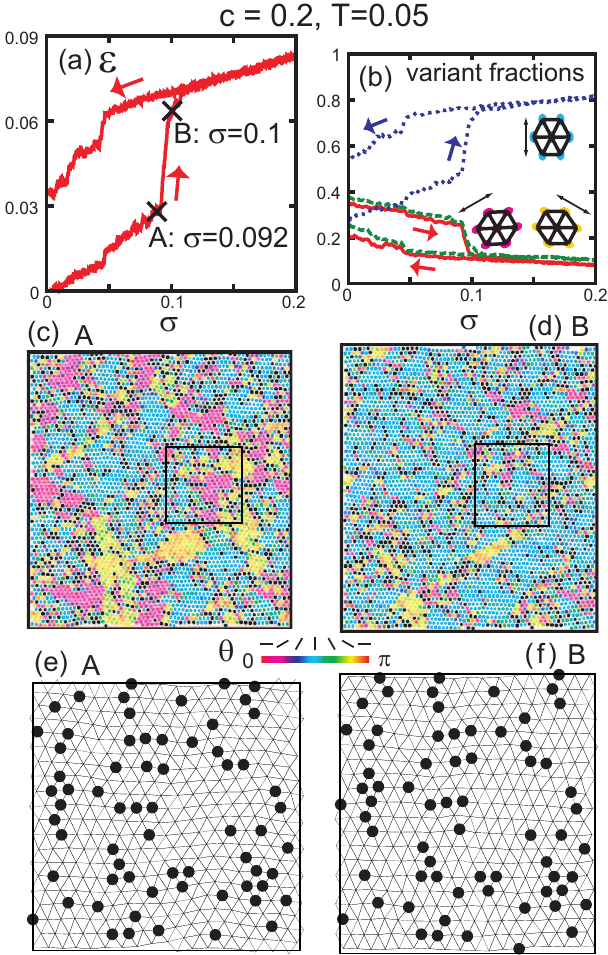}
\caption{Shape memory effect under   stretching  
 for $c=0.2$ and   
 $T=0.05$. (a) Strain $\ve$ vs applied stress ${\sigma}$. 
Between points A ($\sigma= 0.092$)  and B ($\sigma= 0.10$), 
the variant  elongated along the $y$ axis 
increases yielding  soft elasticity.
The  favored variant remains 
dominant on the return path. 
(b)  Fractions of the three variants during  the cycle, 
which are elongated  along  the three crystal axes. 
In (c) and (d),  $[\theta_i]_\pi$ are  written 
at points A and B in (a). 
In e) and (f),   Delaunay diagrams of the box regions 
in (c) and (d) are given.
}
\end{center}
\end{figure}

\begin{figure}[t]
\begin{center}
\includegraphics[width=240pt]{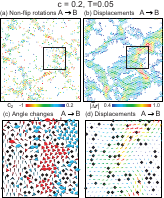}
\caption{(a) Ellipses  with large non-flip rotation 
 $c_{2i}<0.2$  and (b) 
particles with large non-affine displacement 
 $|\tilde{\Delta} {\bi r}_i|>0.4$ between two points A 
and B in Fig.12, where soft elasticity is realized. Colors are given 
according to the color bars. 
(c) Expanded snapshot of the ellipses in  the box region in (a), 
 where each ellipse $i$ is written as an circular sector 
with two arcs parallel to  ${\bi n}_i(t_0)$ 
and   ${\bi n}_i(t_0+t)$ written in blue  for clockwise rotation 
and in red  for counterclockwise rotation. (d) 
Expanded snapshot of the particles with large non-affine  displacement 
 in  the box region in (b). 
}
\end{center}
\end{figure}

\begin{figure}[htbp]
\begin{center}
\includegraphics[width=220pt]{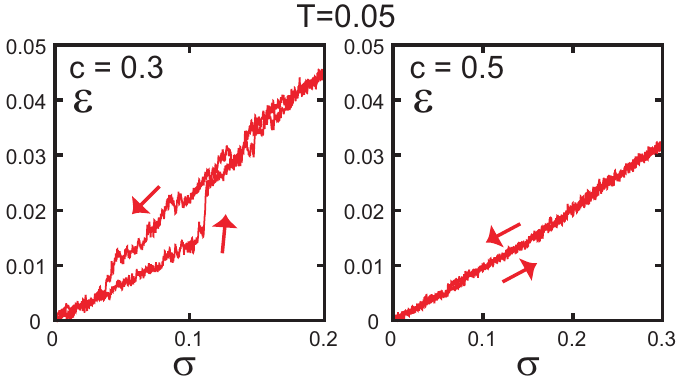}
\caption{Stress-strain curve for $c=0.3$ (left) 
and 0.5 (right).  For $c=0.3$ the loop 
is closed at positive  $\sigma$ and 
there remains  no remnant strain (superelasticity). 
For $c=0.5$ there is no hysteresis and  curve is 
linear in the range  $0\le \sigma \le 0.3$. 
}
\end{center}
\end{figure}
\begin{figure}[htbp]
\begin{center}
\includegraphics[width=220pt]{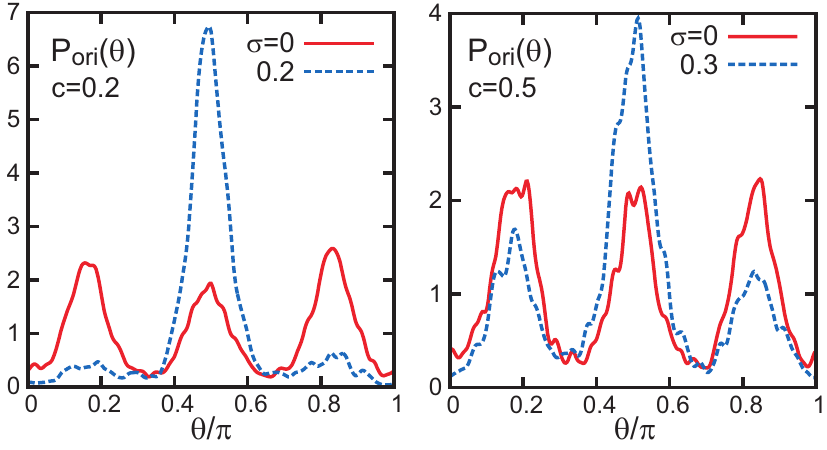}
\caption{Orientational distribution $P_{\rm ori}(\theta) $ 
in Eq.(5.9) for $c=0.2$ (left) 
and 0.5 (right) for the initial point 
$\sigma=0$  and the point with  maximum 
$\sigma$ ($0.2$ or 0.3) 
in the stress cycle. It has three peaks 
with equal heights for $\sigma=0$ and 
the peak at $\theta \cong \pi/2$ is increased 
at the maximum $\sigma$.  
}
\end{center}
\end{figure}

\subsection{Shape memory effect}

We stretched the system along the $y$ axis 
keeping the cell shape rectangular 
under the isothermal condition    at $T=0.05$. 
We mention similar simulations in Refs.\cite{Falk,Suzuki}.  
In the following figures, the  $x$ and $y$ axes are 
 in the horizontal and vertical directions, 
respectively.  One crystal axis of the crystal 
was made parallel to the $x$ axis.  We controlled the space  average of  
the $yy$  component of the stress. Its value is written as    
\be 
\sigma= \av{\sigma_{yy}}_{\rm s}, 
\en 
where  $\av{\cdots}_{\rm s}$ 
denotes the space average. 
 The  system was assumed to be stress-free along the $x$ axis. Thus,  
\be 
\av{\sigma_{xx}}_{\rm s}=0,  
\en  
which is possible owing  to  the attractive 
part of  the  potential \cite{Falk}. 
From the symmetry of the geometry, 
we also have $\av{\sigma_{xy}}_{\rm s}=0$.

We started from  a stress-free state with $\sigma=0$ 
in a square cell with length $L_{0}$. 
With applied stress, the cell lengths   along the $x$ and $y$ axes 
are  changed to $L_x$ and $L_y$. 
The average strain  along the $y$ axis  is written as   
\be 
\varepsilon= L_y/L_{0}-1 . 
\en 
The  effective Young modulus is defined by  
\be 
{E_e}=  ({d\ve}/{d\sigma})_T^{-1}.
\en   
This means that 
an  incremental change $\delta\sigma$ in the applied stress 
gives rise to an incremental  strain  $\delta\ve= E_e^{-1}\delta\sigma$.  
Though  $E_e$ nonlinearly depends on $\sigma$ in our case, 
we  may also define  the effective shear modulus $\mu_e$ 
using the linear elasticity relation   $E_e= 4K\mu_e/(K+\mu_e)$, 
 where $K$ is the bulk modulus 
related to the volume $V$ by 
$2\sigma/K= V/L_{y0}^2-1$. In our system, 
  $K\gg E_e/4$ holds  so that  
\be
\mu_e=E_e/(4-E_e/K)\cong E_e/4. 
\en
Hereafter, we measure   $\sigma$ and  $E_e$  in units of 
$\epsilon/\sigma_1^2$. 


In Fig.12(a), we increased $\sigma$ slowly as  
$d\sigma/dt=4\times 10^{-6}$ 
from 0 up to 0.2 and then decreased $\sigma$ back to 
0 as $d\sigma/dt=-4\times 10^{-6}$. 
The stretching pass 
is divided into four parts: (1) $0<\sigma<0.04$,  (2)  $0,04<\sigma<0.09$, 
(3) $0.09<\sigma<0.12$, and (4)  $0.12<\sigma<0.20$, 
while the return path is divided into two parts: 
(5) $0.20>\sigma>0.05$, and (6)  $0.05>\sigma>0$. 
The effective Young modulus $E_e$ is very  small 
in the range (3). In fact,  we have $E_e \cong 0.25$ between 
 two points A ($\sigma= 0.092$)  and B ($\sigma= 0.10$) 
in Fig.12(a),  while $E_e\cong 4$ in the range (1) 
and   $E_e\cong 8$ in the ranges (4) and (5).  
In addition, there appears a remnant strain ($\sim 0.03$) 
at the final point $\sigma=0$. Furthermore, 
if $T$ was  raised above 0.2 in this  final 
state, an orientationally disordered state was realized    
and a square shape of  the cell was restored.

In Fig.12(b), we show the volume fractions of the three variants. 
See the sentences below Eq.(5.9) as to how they can be 
determined. We can see that the favored domains  
elongated along the $y$ axis increase and the disfavored 
ones decrease upon stretching.  However, the favored domains 
do not much decrease  in the return path, giving rise to the  
 remnant strain. We also show the snapshot of the 
orientations at  point A in (c) and that at   point B in (d), 
between which  the fractions of the favored variant 
are considerably  different. In this stress  cycle,  a  
  history-dependent loop is realized, where  the impurities pin 
 the  orientation domains in  quasi-stationary 
states under very slow variations of $\sigma$.   
In (e) and (f), we give   
 Delaunay diagrams of the box regions in (c) and (d), 
where   local strain variations can be seen without defects.

In Fig.13, we show the orientational and positional changes 
between two points A and B in Fig.12, 
which exhibit  conspicuous large-scale heterogeneities.  
Displayed are  (a)  ellipses  with large non-flip rotations with 
\be 
c_{2i}(A,B)= \cos[2\theta_i(A)-2\theta_i(B)]<0.2
\en  
and  (b) particles with large non-affine displacement  
\be 
|\tilde{\Delta} {\bi r}_i(A,B)|>0.4.
\en  
Here, we write the angles and positions 
at A  as $\theta_i(A)$ and ${\bi r}_i(A)=({x}_i(A),{y}_i(A))$ 
and those at B as $\theta_i(B)$ and ${\bi r}_i(B)=({x}_i(B),{y}_i(B))$. 
The $x$ and $y$  components of $\tilde{\Delta} {\bi r}_i(A,B)$ 
are defined by 
\bea 
&&\hspace{-5mm} 
\tilde{\Delta} {x}_i(A,B)= {x}_i(B)-  x_i(A) L_x(B)/L_x (A),\nonumber\\
&&\hspace{-5mm}
\tilde{\Delta} {y}_i(A,B)= {y}_i(B)- y_i(A)L_y(B)/L_y (A),
\ena  
where $L_\mu(A)$ and $L_\mu(B)$ are  the cell 
lengths at A and B ($\mu=x,y$).  For affine deformations, 
 $\tilde{\Delta} {\bi r}_i(A,B)$ vanishes. 

As in Fig.11(c), Fig.13(c) displays 
the ellipses in the box region in (a)  
written as circular sectors, whose 
arcs are parallel to  ${\bi n}_i(A)$ 
and   ${\bi n}_i(B)$ (in blue  for $\theta_i(A)<\theta_i(B)$ 
and in red  for $\theta_i(A)>\theta_i(B)$). The rotations are more collective 
with weaker thermal fluctuations  than in Fig.11(c). 
In (d), the particles with large non-affine displacement   (5.8) 
are written as arrows, which also indicates collective 
motions upon  stretching. 
In both  (c) and (d), the heterogeneities  are strongly 
correlated with the inhomogeneous impurity distribution. 
The simulation time between A and B 
is 2000, so  the flip numbers $n_i(A,B)$ between A and B are 
small. In fact,  the ellipse number without flips $n_i(A,B)=0$ 
is $0.75N_1$, while that with  $n_i(A,B)\ge 5$ is $0.04N_1$ with  
$n_{\rm max}(A,B)=10$.

We also find that  the shape memory effect 
becomes weaker with increasing $c$, where 
the  domain size is decreased. 
 In Fig.14, the hysteresis loop diminishes 
 for $c=0.3$ and  vanishes for $c=0.5$. 
We recognize that 
mesoscopic orientational order is responsible for 
 the singular mechanical response. 
However, a   unique feature arises 
for $c=0.3$, though the loop is smaller. That is,   the loop is closed 
at $\sigma\sim 0.03$ on the return path and  the initial and final 
points  coincide,  
resulting in no remnant strain at $\sigma=0$. 
It is worth noting that 
this  stress-strain behavior, called superelasticity,    
 has been  observed in  metallic alloys  as 
 a stress-induced martensitic 
 phase transition \cite{marten,Ren,RenReview}.
In TiNi, this   superelasticity effect appears at higher temperatures 
 than the shape memory effect.  
For a  model alloy system, 
Ding {\it et al.} \cite{Suzuki} numerically studied  
the superelasticity effect. 

Furthermore, in Fig.15, we plot the angle distribution, 
\be 
P_{\rm ori}(\theta)= 
\sum_{i \in 1}\delta([\theta_i]_\pi-\theta)/N_1,     
\en 
at  $T=0.05$. Each curve  was results of  a single run. 
This distribution has three peaks both for $c=0.2$ and $0.5$ 
in the directions of the three crystal axes and  
the peak at $\theta\cong \pi/2$  increases  
after stretching. This behavior is consistent with 
Fig.12(b). Here, we  divide the ellipses into the 
three groups with $n/3\le [\theta_i]_\pi/\pi<(n+1)/3$ 
 ($n=0,1,2$)  and  calculate  their  volume fractions  
during stretching.

\begin{table}
\caption{Stress change $\Delta\sigma$, strain change $\Delta\ve$, 
Young modulus $E_e$, change of fraction of the favored variant 
$\Delta\phi_{\rm va}$, 
fraction of orientational strain change $\gamma_{\rm ori}$, 
and ratio ${E_e}/\gamma_{\rm el}$ for $c=0.2$ and $T=0.05$, which are 
calculated  between two points on 
the stress-strain curves. 
 The last quantity should be equal to $E_0$ 
from Eq.(5.12). 
}
\begin{tabular}{|c|c||cccccc|}
\hline
$c$ &
$\sigma$ & $\Delta\sigma$ &
$\Delta\ve$ &
$E_e$ &
$\Delta\phi_{\rm va}$ &
$\gamma_{\rm ori}$ &
${E_e}/\gamma_{\rm el}$ \\
\hline
0.2 & 0 $\to$ 0.04 &0.04& 0.010 & 4 & 0.05 & 0.75 & 16 \\
\hline
0.2 & 0.092 $\to$ 0.1 &0.008&  0.032 & 0.25 & 0.21 & 0.98 & 15 \\
\hline
0.2 & 0.12 $\to$ 0.2 &0.08&  0.011 & 8 & 0.04 & 0.55 & 18 \\
\hline
\hline
0.3 & 0 $\to$ 0.1 &0.1& 0.014 & 7 & 0.06 & 0.64 & 18 \\
\hline
0.3 & 0.11 $\to$ 0.12 &0.11& 0.010 & 1 & 0.06 & 0.94 & 18 \\
\hline
0.3 & 0.12 $\to$ 0.2 &0.08& 0.018 & 4 & 0.09 & 0.72 & 16 \\
\hline
\hline
0.5 & 0 $\to$ 0.15 &0.15& 0.014 & 11 & 0.04 & 0.43 & 19 \\
\hline
0.5 & 0.15 $\to$ 0.3 &0.15& 0.018 & 8 & 0.07 & 0.56 & 18 \\
\hline
\end{tabular}
\end{table}

\subsection{Orientational strain} 
  
On the stress-strain curve, we consider two  points 
between which the curve  is nearly linear. 
See  Figs.12 and 13 for examples.   
 From   Eq.(5.4) the stress change  $\Delta \sigma$ and 
 the strain change $\Delta \ve$ are related by  
\be 
\Delta \ve= E_e^{-1}  \Delta \sigma, 
\en  
in terms of the effective Young modulus $E_e$. 
Generally, in the presence of a (proper)  coupling 
between   strain  and  orientation, $\Delta \ve$   
consists of  three  parts as   
\be  
\Delta \ve =\Delta \ve_{\rm el} + \Delta \ve_{\rm pl}+ \Delta \ve_{\rm ori} ,
\en 
First, the elastic part $\Delta \ve_{\rm el}$ 
is approximately related to  $\Delta\sigma$ by 
the linear elasticity relation, 
\be  
\Delta \ve_{\rm el}= E_0^{-1}  \Delta \sigma ,
\en  
where $E_0 $ is the (bare) Young modulus in a single variant 
state  being of order 20 in our case.  
 Second, the plastic part $\Delta \ve_{\rm pl}$ 
is due to plastic deformations. In the present example, 
there is  no defect generation up to a large 
applied stress $\sigma_{\rm y}$, where  $\sigma_{\rm y} 
\sim 0.5$ for $c=0.2$. 
Thus,  we  set $\Delta \ve_{\rm pl}=0$ in this paper.  
 Third, the orientational part  $\Delta \ve_{\rm ori}$ is 
related to the change of the volume fraction of the 
favored variant $\Delta\phi_{\rm va}$ as 
 \be 
\Delta \ve_{\rm ori} = A_0 \Delta\phi_{\rm va}. 
\en  
where we calculate 
$\Delta\phi_{\rm va}$ from the angle distribution $P_{\rm ori}(\theta)$ 
in Eq.(5.9) (see Fig.15). We may determine the coefficient $A_0$ 
if we apply the relation (5.13) 
between the  initial point $(\sigma,\ve)=(0,0)$ 
and the final point $(\sigma,\ve)=(0,0.03)$ 
on the stress-strain loop in Fig.12(a). 
Using  $\Delta\phi_{\rm va}=0.2$ between these two 
points, we find   $A_0= 0.15$ for $c=0.2$.

It is convenient to  define the ratios, 
\be 
\gamma_{\rm el}=\Delta \ve_{\rm el}/\Delta \ve,\quad 
\gamma_{\rm ori}=\Delta \ve_{\rm ori}/\Delta \ve,
\en 
In this paper, we have 
$\gamma_{\rm el}+\gamma_{\rm ori}=1$ from $\Delta \ve_{\rm pl}=0$.
From Eq.(5.10) we obtain 
\be 
E_e=E_0(1-\gamma_{\rm ori}) =E_0\gamma_{\rm el}.
\en 
In Table 1, we give 
examples of the quantities, 
 $\Delta\sigma$,  $\Delta\ve$, 
 $E_e$, $\Delta\phi_{\rm va}$, 
 $\gamma_{\rm ori}$, and 
${E_e}/\gamma_{\rm el}$ for $c=0.2$, 0.3, and 0.5 
at $T=0.05$. We set $A_0=0.15$ for these three 
concentrations, though it has been obtained for 
$c=0.2$.  From Eq.(5.13) 
the last quantity ${E_e}/\gamma_{\rm el}$ should be equal to $E_0$ 
and is indeed calculated to be around 18.

\setcounter{equation}{0}
\section{Summary and remarks}

We have presented an   angle-dependent 
Lennard-Jones potential for binary mixtures in Eqs.(2.3) and (2.5) 
and performed MD  simulation in two dimensions 
varying  the concentration $c$ and the temperature $T$. 
In this paper, the aspect ratio 
is 1.2 and the size ratio is 1.2, so  the crystal 
order is   realized in all the examples treated. 
Our main results are as follows.\\
(i) In Sec.II, we have  presented 
our model potential, which has  the quadrupolar symmetry 
and  is characterized by  the 
anisotropy strength of repulsion $\chi$  in Eq.(2.5).  
\\ 
 (ii)In Sec.III, we have presented  results  
of $NVT$ simulation. First, the orientation amplitude $\av{S}$ 
and the constant-volume specific heat $C_V$ 
have been presented as functions of $T$ around 
the orientational transition in Fig.1. 
Second, frozen orientational configurations at $T=0.05$ 
have been displayed for $c=0$ in Fig.2 and 
for six concentrations  in Fig.3. 
These snapshots indicate how the domains 
are fragmented with increasing $c$ from 
martensitic multi-domain states to quadrupolar 
 glassy  states.   In our system, 
the circular  impurities exhibit significant clustering and 
anchor the ellipses around them  
in the planar  alignment, as illustrated in Fig.4. 
  \\  
(iii) In Sec.IV, we have studied the rotational dynamics 
of the ellipses, where the orientation  relaxations are  two-fold. 
In Fig.5,  $G_1(t)$ decays 
due to the thermally activated  flip rotations 
even  at low $T$, while  $G_2(t)$ decays due to 
 the non-flip ones  appreciable only in plastic crystal. 
The  corresponding relaxation times $\tau_1$ and $\tau_2$ are much separated 
with $\tau_1\ll \tau_2$.  The   distribution 
of angle changes   $G(t,\theta)$ in Eq.(4.5)  
 evolves  as in Fig.6 due to flip rotations 
and may be approximated as in Eq.(4.8) at low $T$. 
The flip numbers $n_i$  in an appropriate 
 time interval with width $t_f$ have been calculated to 
give the broad flip distribution 
in Eq.(4.21)  and in Fig.9. The rotational diffusion constant $D_R$ 
from the angular mean-square displacement   
is determined by rapidly flipping ellipses as in  Eq.(4.22), 
while $G_1(t)$ is determined by 
 those without  flips as in Fig.9(d). 
The flip activity is closely correlated 
with the impurity 
distribution and is heterogeneous  due to 
the impurity clustering as  in Fig.10.  
\\ 
(iv) In Sec.V, we have examined rheology of  
orientational   glass at   $T=0.05$. For $c=0.2$, we have found a   
 shape memory effect  due to the orientation-strain  coupling in Fig.12, 
where  the stress-strain loop  ends at zero stress 
with a remnant strain. When soft elasticity  appears, 
 the angle changes  and the non-affine displacements become 
 highly heterogeneous and collective as in Fig.13. 
 For $c=0.3$,  we have found a  superelasticity effect in Fig.14, 
where the  loop is closed at nonvanishing stress. 
The angle distribution $P_{\rm ori}(\theta)$ in Eq.(5.9) 
has three peaks and is changed by applied stress as in Fig.15.  
For an incremental stress  change (superimposed on a main  stress) 
 gives rise to a small elastic strain and a small orientational strain. 
The effective Young modulus $E_e$ is decreased due to 
the appearance of the orientational strain. 
In Table 1, we have calculated these quantities on the stress-strain 
curves for $c=0.2$, 0.3, and 0.5.  
\\

We further make critical remarks as follows:\\
(1) The aspect ratio in this paper is rather close to unity. 
We should examine the glass transitions 
for various aspect ratios and molecular shapes. 
For large anisotropy,  
liquid crystal  phases should appear \cite{Frenkel1}, 
where the impurity   effect  has not yet 
well understood. Mixtures of two species of anisotropic particles 
should also  be studied.\\
(2) Though not well recognized, 
the impurity clustering can    be crucial  
in various glassy systems. As in this paper, 
the   impurity distribution should  
influence the underlying phase transition 
and     dynamics of some order parameter. 
We need systematic experiments on various 
kinds of mesoscopic 
heterogeneities in glass 
for a wide range of impurity 
concentrations. \\ 
(3) In real systems, the  quadrupolar  
behavior can be expected only  when the constituent 
 molecules  carry  small dipole  moments 
and exhibit no head-to-tail order at low $T$ 
\cite{ori}.  Experimentally,  the dipolar freezing in mixtures of KCN-KBr 
(slowing down of reorientational motions of CN$^{-}$)  was found to occur 
 at low $T$ in  a quadrupolar glass state \cite{qua}.  
 On the other hand, for 
molecules with large dipole moments, 
 applied electric field  can be important 
 \cite{apply} and a ferroelectric transition can even occur.  \\
(4) As mentioned in Sec.I, one-component  systems of 
globular molecules  become 
orientational glass \cite{Seki,Yamamuro,C60}. 
However, to understand  this phenomenon, 
we cannot use the physical picture for mixtures  in this paper.\\
(5) In our recent  paper \cite{Takae-double},
 we have  examined  the effect  of  small 
  impurities, to which host anisotropic particles  
are homeotropically anchored \cite{PG}. 
We stress that there can be a variety of 
angle-dependent  molecular interactions, giving rise to 
a wide range  of rotational and translational 
glass formers.\\

\begin{acknowledgments}
This work was supported by Grant-in-Aid 
for Scientific Research  from the Ministry of Education, 
Culture,  Sports, Science and Technology of Japan. 
The authors would like to thank Takeshi Kawasaki  and  
Osamu Yamamuro    for valuable discussions. 
K. T. was supported by the Japan Society for Promotion of Science.
The numerical calculations were carried out on SR16000 at YITP in Kyoto
University.
\end{acknowledgments}

\vspace{2mm}
\noindent{\bf Appendix A: 
Orientational  order parameter}\\
\setcounter{equation}{0}
\renewcommand{\theequation}{A\arabic{equation}}

Here, we    introduce an   
orientation tensor ${\aQ}_i =
 \{ Q_{i\mu\nu}\}$ ($\mu,\nu=x,y$) for each anisotropic particle $i\in 1$ by  
\be
\aQ_i=
\frac{1}{1+n_{\rm b}^i} \bigg({\bi n}_i{\bi n}_i
 + \sum_{j\in {\rm bonded}} {\bi n}_j {\bi  n}_j\bigg) 
-\frac{1}{2}\aI,  
\en 
where ${\aI}= \{\delta_{\mu\nu}\}$ is the unit tensor. 
In the summation over  $j$, 
we pick up  the ellipses  
in the region   $|{\bi r}_{ij}| <1.3$ ($i,j \in 1$). The   
 $n_{\rm b}^i$ is   the number of these bonded ellipses. 
Thus, this tensor is a coarse-grained 
orientational order parameter as in  liquid crystal systems.  
If a  hexagonal 
 lattice is formed, 
 the   nearest neighbor particles are 
included in this definition, 
so $n_{\rm b}^i \sim 6$. This $2 \times 2$  tensor 
  is traceless and symmetric, so it  
 may   be expressed  
 as 
\be 
Q_{i\mu\nu}=  \sqrt{S_i} 
({ d}_{i\mu} {  d}_{i\nu} -\delta_{\mu\nu}/2),
\en 
in terms of  an   amplitude $S_i$ and 
 a  unit vector (director)   ${\bi d}_i=(d_{ix}, d_{iy})$.  
For each $i$,  $S_i$ may be expressed as    
\be 
{S_i}= 2 \sum_{\mu,\nu}  Q_{i\mu\nu}^2,  
\en 
which  increases  up to unity in ordered regions  
at low $T$ and is about   $ 0.1$  in disordered crystals   
 due to the thermal fluctuations.  The  degree of overall orientational  order 
is  represented by    the  average,  
\be 
\av{S}=\frac{1}{N_1} \sum_{ i \in 1} S_i= \frac{2}{N_1} 
 \sum_{i\in 1}\sum_{\mu,\nu}  Q_{i\mu\nu}^2.   
\en 
See Fig.1 for a plot  of  $\av{S}$ vs $T$.

\vspace{2mm}
\noindent{\bf Appendix B: 
Flip times and numbers }\\
\setcounter{equation}{0}
\renewcommand{\theequation}{B\arabic{equation}}

Here, we determine  a series of  flip times,  
$t_0+t_{i1}<t_0+ t_{i2}<t_0+ t_{i3}<\cdots$,  for each ellipse $i$ in 
time interval $[t_0,t_0+t_f]$, where $t_f\gg \tau_1$. 
(i) The first flip time $t_0+t_{i1}$ 
is determined in terms of  
 $\Delta\theta_i (t)
=\Delta\theta_i (t_0,t+t_0)$   by 
\be 
|\Delta\theta_i(t_{i1}) |= 2\pi/3.
\en  
  For $t>t_{i1}$ we introduce  a shifted angle change, 
\be 
\Delta\theta_{i1} (t) = 
\Delta\theta_i(t) \pm \pi,
\en 
 where $+\pi$ or $-\pi$ 
is chosen such that $|\Delta\theta_{i1}(t_{i1}+0)| 
<\pi/2$.  
(ii) The second flip time  $t_0+ t_{i2}$ is determined by  
\be 
|\Delta\theta_{i1}(t_{i2})|=2\pi/3.
\en  
For $t>t_{i2}$, we again shift the angle change as   
\be 
\Delta\theta_{i2} (t)=\Delta\theta_{i1}(t)  \pm \pi,
\en  
 where 
 $|\Delta\theta_{i2}(t_{i2}+0)| < \pi/2$. (iii)  Repeating  these 
procedures yields the successive flip times. See  Fig.8.   
Within  any time interval $[t_0, t_0+t_f]$, 
each ellipse  flips at   times 
$t_0+t_{i1},t_0+t_{i2}, \cdots, t_0+ t_{in_i}$ ($t_{i1}>0$ and  
$t_{in_i}<t_f$), with  $n_i$ being  
the flip number of ellipse $i$.


\begin{thebibliography}{0}



\bibitem{ori} 
U. T. H\"{o}chli, 
K. Knorr, and A. Loidl, Adv. Phys. {\bf 39}, 405 (1990).
\bibitem{Binder} K. Binder, and J.D. Reger, Adv. Phys. {\bf 41}, 547 (1992).

\bibitem{Kob-Binder} 
K. Binder and W. Kob, {\it 
Glassy Materials and Disordered Solids} 
(World Scientific, Singapore, 2005).



\bibitem{Sherwood} 
{\it 
The Plastically crystalline state: 
orientationally disordered crystals},   
edited by John N. Sherwood  
(John  Wiley $\&$ Sons, Chichester, 1979). 


\bibitem{St} K. Knorr and A. Loidl,  Phys. Rev. B {\bf 31},
 5387 (1985).
 

\bibitem{Sullivan} N. S. Sullivan, M. Devoret, B. P. Cowan, and C. Urbina, 
Phys. Rev. B {\bf 17}, 5016 (1978).


 \bibitem{Mertz}
B. Mertz  and A. Loidl, 
EPL {\bf 4},  583 (1987). 


\bibitem{sound} 
K. Knorr, U. G. Volkmann, and A. Loidl, 
Phys. Rev. Lett. {\bf 57}, 2544 (1986).


\bibitem{Raedt} K. H. Michel and J. Naudts, J. Chem. Phys. {\bf 67}, 
547 (1977); B. De Raedt,  K. Binder, and K. H. Michel, 
J. Chem. Phys. {\bf 75}, 2977 (1981).

\bibitem{Knorr} 
K. Knorr, Phys.Rev.B {\bf 41}, 3158 (1990).

\bibitem{Bell} R. M. Lynden-Bell and K. H. Michel, 
Rev, Mod. Phys. {\bf 66}, 721 (1994).

\bibitem{Harris}  A.B. Harris, Physica A {\bf 205}, 154 (1994). 
\bibitem{Klein} 
L. J. Lewis and M. L. Klein, Phys. Rev. B {\bf 40}, 4877 (1989); 
ibid. {\bf 40},  7080 (1989).



\bibitem{Seki} K. Adachi, H. Suga, and S. Seki,
 Bull. Chemi. Soc.Japan {\bf 41}, 1073 (1968).
\bibitem{Yamamuro} O. Yamamuro, M.Ishikawa, 
I. Kishimoto, J.J. Pinvidic, and T.Matsuo,
 J. Phys. Soc. Japan {\bf 68} 2969 (1999); 
O. Yamamuro, H. Yamasaki, Y. Madokoro,
I. Tsukushi,  and T.  Matsuo,  J. Phys.: Condens. Matter 
 {\bf 15}, 5439 (2003). 

\bibitem{C60}
F. Gugenberger, R. Heid, C. Meingast, P. Adelmann, M. Braun, 
H. W{\"u}hl, M. Haluska,  and H. Kuzmany, 
Phys. Rev. Lett. {\bf 69}, 3774 (1992).



 
\bibitem{Kob1} S. K{\"a}mmerer, W. Kob, and R.  Schilling, 
Phys. Rev. E {\bf 56},  5450 (1997); 
C. De Michele and D. Leporini, Phys. Rev. E {\bf 63}, 
 036702 (2001);
S.-H. Chong, A. J. Moreno, F. Sciortino, and W. Kob,
Phys. Rev. Lett. {\bf 94}, 215701 (2005); 
A. J. Moreno, S.-H. Chong, W. Kob, and F. Sciortino, 
J. Chem. Phys. {\bf 123}, 204505 (2005); 
S.-H. Chong and W. Kob,
Phys. Rev. Lett.  {\bf  102}, 025702 (2009). 


\bibitem{Lewis} L. J. Lewis and G. Wahnstr{\"o}m, 
 Phys. Rev. E {\bf 50}, 3865 (1994);
J. Non-Cryst. Solids {\bf 172}, 69 (1994);  
T. G. Lombardo, P. G. Debenedetti, and F. H. Stillinger,
J. Chem. Phys. {\bf 125}, 174507 (2006); 
M. G. Mazza, N. Giovambattista,  F. W. Starr, and H. E. Stanley,
Phys. Rev. Lett.{\bf 96}, 057803 (2006).
N. B. Caballero, M. Zuriaga, M. Carignano, and P.  Serra, 
J. Chem. Phys. {\bf 136}, 094515 (2012).

\bibitem{Schilling} 
P. Pfleiderer, K. Milinkovic, 
and T. Schilling, EPL {\bf 84},  16003 (2008).


\bibitem{Sch} R. Zhang and K. S. Schweizer, 
J. Chem. Phys. {\bf 133}, 104902 (2010); 
ibid. {\bf 136}, 154902 (2012).


\bibitem{EPL} K.  Takae and A.  Onuki, 
EPL  {\bf 100},  16006 (2012).

\bibitem{Hamanaka} T. Hamanaka and A. Onuki, 
Phys. Rev. E {\bf 74}, 011506 (2006);  ibid. 
 {\bf 75}, 041503 (2007).


\bibitem{Tanaka} 
H. Tanaka, T. Kawasaki, and H. Shintani, and K. Watanabe, Nature Mater.
{\bf 9}, 324 (2010).
\bibitem{Takae-double} K.  Takae and A.  Onuki, accepted for 
publication (PRE)  (arXiv:1309.0779). 
\bibitem{Cowley} R.A. Cowley, Phys. Rev. {\bf B13}, 4877 (1976).  
\bibitem{Onukibook}
A. Onuki, 
\textit{Phase Transition Dynamics} (Cambridge University Press, 
Cambridge, 2002).

\bibitem{marten} 
H. Warlimont and L. Delaey, Progr. Mater. Sci. {\bf 18}, 
1 (1974).  


\bibitem{RenReview}
K. Otsuka and X. Ren, Prog. Mater. Sci. {\bf 50}, 511 
(2005). 

\bibitem{Ren}
S. Sarkar,  X. Ren, and K. Otsuka, 
Phys. Rev. Lett. {\bf 95}, 205702 (2005); 
Y. Wang, X. Ren, and K. Otsuka,  
Phys. Rev. Lett. {\bf 97}, 225703 (2006).


\bibitem{Suzuki} X. Ding, T. Suzuki, X. Ren, J. Sun,
 and K. Otsuka, Phys. Rev. B {\bf 74},  104111 (2006). 

\bibitem{PGgel} P. G. de Gennes,   
C. R. Acad. Sci., Ser. B {\bf 281}, 101 (1975); 
M. H\'ebert, R. Kant, and P. G. de Gennes, J. Phys. I France {\bf 7}, 909 (1997).

\bibitem{Tere} 
M. Warner and E. M. Terentjev, {\it Liquid crystal elastomers}
(Cambridge University Press, Cambridge, 2003).

\bibitem{Uchida} N. Uchida,  Phys. Rev. E {\bf  62},  5119 (2000). 

\bibitem{Kai} 
J. K{\"u}pfer and H. Finkelmann, Macromol. Chem. Phys. {\bf 195},
1353 (1994);  Y. Yusuf, 
J.-H. Huh, P. E. Cladis, H.R. Brand, 
H.  Finkelmann, and S. Kai, 
Phys.Rev.E {\bf 71}, 061702 (2005); 
K.  Urayama,  E. Kohmon, M. Kojima, and T. Takigawa, 
Macromolecules {\bf 42}, 4084 (2009). 



\bibitem{Gay}
J. G. Gay and B. J. Berne, J. Chem. Phys. {\bf 74}, 3316 (1981); 
J. T. Brown,  M. P. Allen, E. Martin del Rio, and E. de Miguel, 
Phys. Rev. E {\bf 57},  6685 (1998).






\bibitem{nose}
S. Nos\'e, Mol. Phys. {\bf 52}, 255 (1984). 
\bibitem{Rahman} 
M. Parrinello and A. Rahman,
J. Appl. Phys. {\bf 52}, 7182 (1981).

\bibitem{Frenkel1} 
D. Frenkel and B. M. Mulder, Mol. Phys. {\bf 55}, 1171 (1985); 
 P. Bolhuis and D.  Frenkel,  
J. Chem. Phys. {\bf 106}, 666 (1997);   
C. Vega  and P. A.  Monson, 
J. Chem. Phys. {\bf 107}, 2696 (1997); 
C. De Michele, R. Schilling, and F. Sciortino, 
Phys. Rev. Lett. {\bf 98}, 265702 (2007); 
M. Radu, P. Pfleiderer, and T. Schilling, 
J. Chem. Phys. {\bf 131}, 164513 (2009);  
 M. Murat  and Y. Kantor, Phys. Rev.
 E {\bf 74}, 031124 (2006).


\bibitem{comment1}  
In $NpT$ simulation, we took data for $c=0$ 
at  $T=T_t+ 10^{-3}n$ ($n=0, \pm 1,
\cdots$)  around the transition  
temperatuture $T_t$  waiting for a time interval 
of $10^5$ at each $T$. 
We found a unique discontinuous change without appreciable 
hysteresis, where the entropy 
change was about $k_BN$.   
Hysteresis appeared for shorter waiting times. 




\bibitem{Kitano} 
R. Sinclair and J. Dutkiewicz, 
Acta Metell. {\bf 25}, 235  (1977);  
C. Manolikas and S. Amelinckx,
Phys. Stat. Sol. (a) {\bf 60}, 607 (1980), 
 {\bf 61}, 179 (1980); 
Y. Kitano, K. Kifune, and 
Y. Komura, J. Phys. (Paris) {\bf 49}, 
C5-201 (1988); 
K. Muraleedharan, D. Banerjee, S. 
Banerjee and S. Lele, 
 Phil.  Mag.  A, {\bf 71}, 1011 (1995). 


\bibitem{Chen}  Y. H. Wen, Y. Wang, and L. Q. Chen, 
 Phil. Mag. A. {\bf 80}, 1967 (2000);    
Y. H. Wen, Y. Wang, L. A. Bendersky, and L. Q. Chen, 
Acta Mater. {\bf 48}, 4125  (2000). 

\bibitem{PG} H. Stark, Phys.Rep. {\bf 351}, 387 (2001).

\bibitem{defects} 
A. H. Marcus and S. A. Rice, Phys. Rev. E {\bf 55},  637 (1997); 
D. A. Vega, C. K. Harrison,  D. E. Angelescu, M.  L. Trawick,  
D. A. Huse,  P. M. Chaikin, and R. A. Register, 
Phys. Rev. E {\bf 71}, 061803 (2005); 
Bo-Jiun Lin and Li-Jen Chen, J. Chem. Phys. {\bf 126}, 
034706 (2007); 
Y. Han, N. Y. Ha,  A. M. Alsayed, and A. G. Yodh
 Phys. Rev. E {\bf 77}, 041406 (2008). 
In these papers, the defect density was found to be  very small 
in the solid phase but increase in the hexatic and liquid 
phases at higher $T$. 



\bibitem{Falk}  N. P. Bailey, J. Schi{\o}tz, 
and K. W. Jacobsen, Phys. Rev. B {\bf 73}, 064108 (2006); 
Y. Shi and M. L. Falk
Phys. Rev. B {\bf 73}, 214201 (2006).



\bibitem{qua} U. G. Volkmann, R. B{\"o}hmer, 
A. Loidl,  K. Knorr, 
U. T. H{\"o}chli, S. Hauss{\"u}hl, 
 Phys. Rev. Lett. {\bf 56}, 1716 (1986). 


\bibitem{apply} K.  Takae and A.  Onuki, 
J. Chem. Phys. {\bf 139}, 124108 (2013).






 










\end{thebibliography}
\end{document}